\pdfoutput=1
\documentclass[11pt,table]{article}

\usepackage[final]{acl}

\usepackage{times}
\usepackage{latexsym}

\usepackage[T1]{fontenc}
\usepackage[utf8]{inputenc}

\usepackage{microtype}

\usepackage{inconsolata}

\usepackage{graphicx}
\usepackage{float}
\usepackage{multirow}
\usepackage{booktabs}
\usepackage{bbding}
\usepackage{enumitem}

\title{OMGM: Orchestrate Multiple Granularities and Modalities for Efficient Multimodal Retrieval}

\vspace{-4mm}
\author{Wei Yang\thanks{Work done during an internship at Microsoft Research Asia.}, Jingjing Fu\thanks{Corresponding author}, Rui Wang, Jinyu Wang, Lei Song, Jiang Bian\\
Microsoft Research Asia \\
\texttt{wyang6621@gmail.com},
\texttt{\{jifu,ruiwa,jinywan,lesong,jiabia\}@microsoft.com}}

\begin{document}
\maketitle
\begin{abstract}

Vision-language retrieval-augmented generation (RAG) has become an effective approach for tackling Knowledge-Based Visual Question Answering (KB-VQA), which requires external knowledge beyond the visual content presented in images. The effectiveness of Vision-language RAG systems hinges on multimodal retrieval, which is inherently challenging due to the diverse modalities and knowledge granularities in both queries and knowledge bases. Existing methods have not fully tapped into the potential interplay between these elements.
We propose a multimodal RAG system featuring a coarse-to-fine, multi-step retrieval that harmonizes multiple granularities and modalities to enhance efficacy. Our system begins with a broad initial search aligning knowledge granularity for cross-modal retrieval, followed by a multimodal fusion reranking to capture the nuanced multimodal information for top entity selection. A text reranker then filters out the most relevant fine-grained section for augmented generation. 
Extensive experiments on the InfoSeek and Encyclopedic-VQA benchmarks show our method achieves state-of-the-art retrieval performance and highly competitive answering results, underscoring its effectiveness in advancing KB-VQA systems.

\end{abstract}

\begin{figure}[t!] 
    \centering 
    \includegraphics[width=0.98\columnwidth]
    {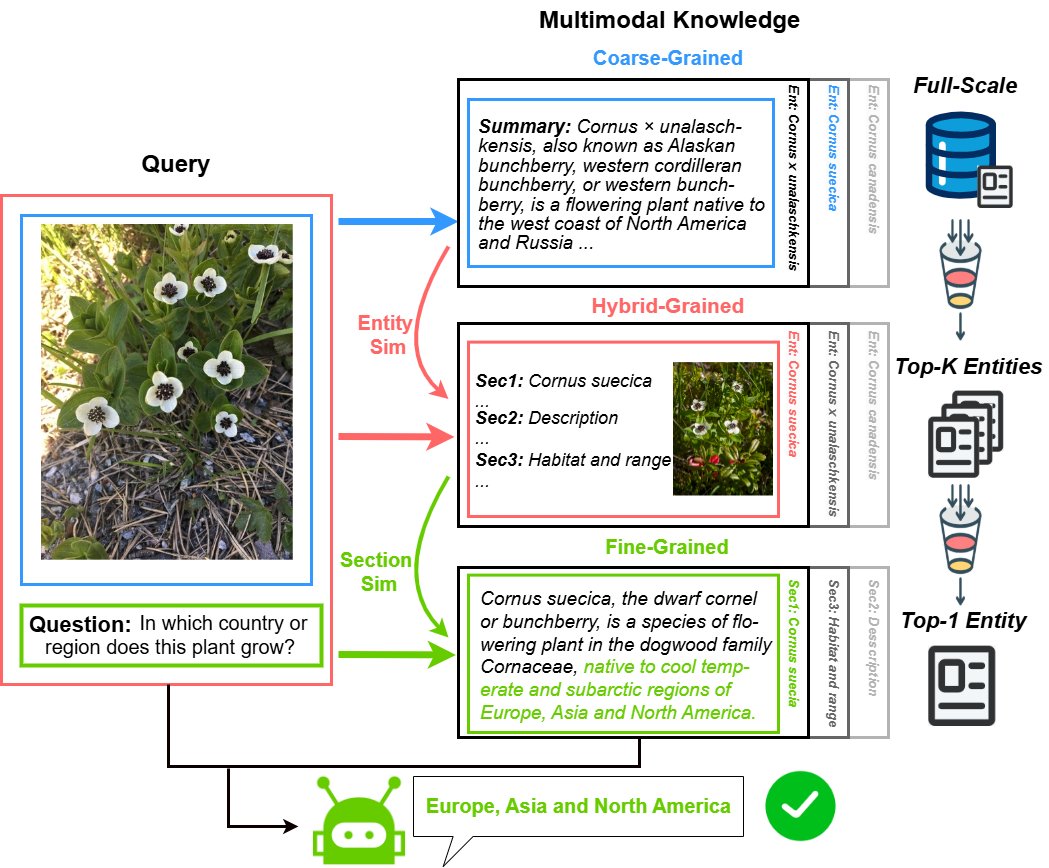} 
    
    \caption{An illustration of our method. First, a coarse-grained cross-modal entity search is performed between entity summaries and the query image to retrieve the top-k entity candidates. Next, a hybrid-grained multimodal fusion reranker uses the multimodal query to retrieve image-section pairs, refining the selection of the most relevant entity. Finally, within knowledge associated with the top-1 entity, fine-grained textual filtering is applied to extract most relevant section, which is used to enhance generation in downstream generator.} 
    \label{fig:intro} 
    \vspace{-6mm}
\end{figure}

\section{Introduction}

Visual Question Answering (VQA) involves answering questions about a given query image by comprehensively understanding its semantic content, demanding proficiency in both visual and textual understanding. Large Language Models (LLMs) have demonstrated remarkable generalization and reasoning capabilities in text-based tasks~\cite{raffel2020exploring, brown2020language}. By integrating visual encoders with LLMs, Multimodal Large Language Models (MLLMs) have emerged as an effective approach for handling VQA tasks, as they can jointly model both image and text representations for enhanced comprehension and reasoning~\cite{alayrac2022flamingo, liu2024improved, li2023blip}. Knowledge-Based Visual Question Answering (KB-VQA) extends this challenge by requiring the incorporation of external world knowledge that transcends the visible elements of the image. In KB-VQA, questions are designed to probe for information pertaining to the image's subject matter, but necessitate insights not directly present within the image itself.

% Visual question answering (VQA) is a task focused on responding to questions related to a given reference image, with a strong emphasis on understanding the semantic content of the image. Large language models (LLMs) possess powerful and general language capabilities. Building on this foundation, multimodal large language models 
%  (MLLMs) employ an image encoder to project images into the language domain, enabling these models to comprehend the semantic content of images. Consequently, the most mainstream and direct approach to handling VQA at present is to leverage MLLMs to perform end-to-end answering by integrating textual questions with image semantics~\cite{li2023blip, liu2024visual}. Unlike traditional VQA, Knowledge-based Visual Question Answering (KB-VQA) involves textual questions that inquire about world knowledge related to the subject of the image, which cannot be directly obtained from the image itself.
% This makes MLLMs, which are not adept at handling long-tail knowledge, prone to encounter hallucination when facing KB-VQA~\cite{mensink2023encyclopedic,chen-etal-2023-pre-trained}.

Retrieval-Augmented Generation (RAG) offers a cost-effective and efficient solution to the challenges of KB-VQA by retrieving query-relevant knowledge from the knowledge base and integrating it as contextual information for response generation~\cite{karpukhin-etal-2020-dense, si-etal-2023-combo}. The effectiveness of this approach depends on efficient retrieval mechanisms capable of identifying the most relevant information from large-scale, heterogeneous knowledge bases. However, multimodal retrieval in KB-VQA introduces complexities beyond the standard text-based retrieval used in most RAG systems. These complexities stem from two key factors: 
\vspace{-2mm}
\begin{itemize}[left=0pt]
\item \textbf{Multiple Modalities.} Both queries and knowledge bases comprise multiple modalities, such as images and text, necessitating diverse relevance assessment strategies that may involve single-modal, cross-modal, or multi-modal approaches. The choice of retrieval schema is influenced by the retrieval model's capabilities and the requirements of specific tasks. 
\vspace{-2mm}
\item \textbf{Hybrid Granularities.} Queries and knowledge bases often involve information at different levels of granularities. For instance, a query might include a query image with coarse-grained knowledge identifying a subject, paired with a question seeking fine-grained details about the subject. Similarly, a knowledge base might consist of articles with coarse-grained overview reference images and titles as well as fine-grained detailed sections containing in-depth information.
\vspace{-2mm}
% c) As The modality and granularity are intertwined, multiple step retrieval are proposed combining retrieval results across multiple steps necessitates the development of advanced design and integration strategies to ensure effectiveness and coherence.The multimodal encoder is not mature enough to generalized solve the retrieval at one step, like text embedding. Fusing the retrieval results from multiple steps requires sophisticated design. 
\end{itemize}   
% modalities and granularity are intertwined within query and knowledge base,     
% However, these methods still have certain limitations: Firstly, KB-VQA’s knowledge-base and queries are complex: 1) They frequently incorporate different types of modality information, such as images and text; 2) They generally exhibit varying levels of granularity—for instance, a query may contain a coarse-grained reference image alongside fine-grained questions; 3) The modalities and granularities are intertwined within the presented information, as seen when an article’s summary provides a coarse-grained overview while individual sections offer fine-grained details.
% A cost-effective and efficient approach to enhance KB-VQA performance is Retrieval-Augmented Generation (RAG), which retrieves knowledge relevant to the query from knowledge base and integrates it as contextual information to support the generation of responses. 
Recent research efforts have explored both single-step and multi-step multimodal retrieval strategies to enhance retrieval effectiveness and ultimately improve answer generation. Single-step approaches retrieve passages directly using a multimodal query~\cite{lin-byrne-2022-retrieval, deng-etal-2025-muka, jian-etal-2024-large,lin-etal-2024-preflmr}. While effective, these methods often require expensive task-specific pretraining and incur high computational costs during inference due to exhaustive full-range searches. Conversely, multi-step retrieval methods adopt hierarchical retrieval strategies that enhance searching efficiency by progressively narrowing the search space. These methods often employ different retrieval modalities at each step, enabling the evaluation of knowledge relevance from multiple perspectives~\cite{caffagni2024wiki, yan-xie-2024-echosight, qi2024rora}. Nevertheless, the potential of multi-step retrieval remains underexplored. Current approaches frequently overlook the intricate interplay between retrieval modalities, granularities, and the sequencing of retrieval steps, limiting their overall effectiveness and adaptability. 
We propose a multimodal RAG system, \textbf{OMGM}, which employs a coarse-to-fine, multi-step retrieval strategy to effectively \textbf{O}rchestrate \textbf{M}ultiple \textbf{G}ranularities and \textbf{M}odalities across queries and knowledge bases, enhancing multimodal retrieval.
% We propose a multimodal RAG system that employs a coarse-to-fine, multi-step retrieval strategy, effectively Orchestrate Multiple Granularities and Modalities in queries and knowledge bases to enhance multimodal retrieval, which we call our system OMGM.
As illustrated in Figure~\ref{fig:intro}, our system operates in three stages: it begins with a coarse-grained cross-modal retrieval to identify an initial pool of entity article candidates, followed by a hybrid-grained multimodal reranker that leverages both coarse-grained and fine-grained knowledge to rerank candidates and select the most relevant entity. Finally, a fine-grained text reranker filters the selected entity's sections to extract the most pertinent sections for augmented response generation.
Throughout the process, query and candidates are aligned based on their granularities, and embedding models are carefully selected to ensure effective multimodal representation. More importantly, these retrieval steps interact sequentially, with similarity scores from earlier stages propagated forward and fused in subsequent steps, enabling a cohesive and context-aware retrieval process.
% To address the aforementioned shortcomings and limitations, we propose a multi-step VQA RAG system as illustrated in Figure~\ref{fig:intro}. Our retrieval system first generates, offline, knowledge summaries that are aligned with the granularity of the reference image, and then uses both to perform a coarse-grained cross-modal searching. Next, among the top-k retrieved entities, we generate multimodal fused features from the query and the entity knowledge’s text and image information with the fine-tuned q-former model, calculate coarse-grained multimodal similarities with late-interaction, and fuse them with the entity similarities from the previous step to rerank the relevant entities. Simultaneously, we compute fine-grained section similarities to be propagated to subsequent step. Finally, within the top-1 reranked entity, we employ a pre-trained text reranker to obtain the fine-grained similarities between the question and the section, which, combined with the previously computed fine-grained similarities, is used to filter out the most relevant fine-grained section to enhance the downstream generator’s performance. 
We conduct extensive retrieval and VQA experiments on two KB-VQA datasets, where our method achieved state-of-the-art retrieval performance and competitive question-answering results compared to other existing multi-step retrieval methods. 

In summary, the main contributions of our work can be summarized as follows:
\begin{itemize}
    \item We propose that multimodal retrieval should be tailored to the characteristics of KB-VQA, specifically in data modality and knowledge granularity. To achieve this, we introduce a coarse-to-fine multi-step retrieval strategy that progressively enhances retrieval quality by leveraging retrieval steps with varying modalities and granularities.
    \item We introduce a trainable multimodal reranker to maximize the utilization of full-modal information while minimizing inference costs by restricting the reranking scope, thereby ensuring both effectiveness and efficiency.
    \item We conducted extensive experiments on the InfoSeek and Encyclopedic-VQA (E-VQA) benchmarks, showcasing the effectiveness of the proposed method. Furthermore, a comprehensive ablation study validates the contribution of each retrieval step, offering valuable insights for designing multimodal retrieval systems in KB-VQA tasks.
\end{itemize}

% (i) We design a multi-step VQA RAG system in which the queries and knowledge at each retrieval step are configured with appropriate modalities and aligned in granularity. Between steps, we propagate similarities to enhance cross-step retrieval interaction. Consequently, this approach optimizes both the granularity and modality arrangement in multimodal retrieval, thereby improving overall retrieval and question answering performance; (ii) Based on the pre-trained q-former model, we fine-tuned a multimodal fusion feature encoder. By encoding both the text and visual content in the queries and candidate knowledge, we can compute multi-granularity similarities, thereby fully and efficiently leveraging multimodal information to achieve enhanced retrieval performance; (iii) We conducted experiments on the InfoSeek and Encyclopedic VQA benchmarks, achieving state-of-the-art retrieval performance and highly competitive question-answering results.
%Introduce the whole pipeline in this paper

%summary three contribution of our work

\section{Related Work}
\subsection{KB-VQA}
Traditional VQA tasks focus primarily on the visual content within images to answer the related textual question~\cite{antol2015vqa}. However, KB-VQA expands this by incorporating external knowledge bases to address questions requiring information beyond the image. Datasets such as OK-VQA and A-OKVQA~\cite{marino2019ok, schwenk2022okvqa} involve visual question answering that require outside knowledge, which consists of general commonsense that lacks distinctiveness. 
 
The emergence of the E-VQA and InfoSeek~\cite{mensink2023encyclopedic,chen-etal-2023-pre-trained} datasets presents a greater challenge for end-to-end LLMs/MLLMs by incorporating varying granularities of encyclopedic knowledge alongside extensive multimodal information. In response to these characteristics of KB-VQA task, our work adopts a multi-step, granularity-aligned retrieval framework, resulting in improved retrieval and VQA performance compared to the current multimodal RAG system~\cite{caffagni2024wiki,lerner2024cross,yan-xie-2024-echosight}.

\subsection{Vision-language RAG}
Retrieval-Augmented Generation (RAG)~\cite{guu2020retrieval} enhances the generation performance of LLMs by retrieving external documents relevant to the input query and using them as guiding context in prompts. In addition to text augmentation, recent work ~\cite{lin-byrne-2022-retrieval,xia-etal-2024-rule} has targeted generation enhancement specifically for vision-language tasks, which are more closely aligned with complex real-world scenarios.

Vision-language retrieval-augmented generation typically entails handling multimodal documents and queries, requiring the retrieval process across these modalities. For example, the approaches introduced in PreFLMR and MuKA~\cite{lin-etal-2024-preflmr, deng-etal-2025-muka} encodes features across various modalities and dimensions, and separately concatenates the retrieval matrices for the query and the candidate, thereby facilitating a fine-grained knowledge search. Wiki-LLaVA~\cite{caffagni2024wiki} and EchoSight~\cite{yan-xie-2024-echosight} employ a hierarchical retrieval strategy, achieving efficient cross-step multimodal retrieval. LLM-RA~\cite{jian-etal-2024-large} and RoRA-VLM~\cite{qi2024rora} adopt LLM-based and similarity-based approaches, respectively, for fine-grained denoising on queries and knowledge, thereby improving  the accuracy of both retrieval and question answering. Additionally, mR$^2$AG~\cite{zhang2024mr} and ReflectiVA~\cite{cocchi2024augmenting} emphasize reflective processing. By fine-tuning large models, they leverage token outputs to drive the retrieval process and subsequently perform relevance-based re-screening and modifications on both the retrieved content and the generated answers. Unlike previous methods, our system constructs a multi‐step retrieval process that integrates granularity alignment with cross‐step joint retrieval. At each step, we align queries with the corpus using techniques such as summary extraction, multimodal fusion, and question‐based section denoising, then fuse results via similarity fusion to boost recall and enhance downstream KB‐VQA performance of MLLMs/LLMs.

 \begin{figure*}[t!]% 选择合适的位置
    \centering
    \includegraphics[width=0.94\linewidth]{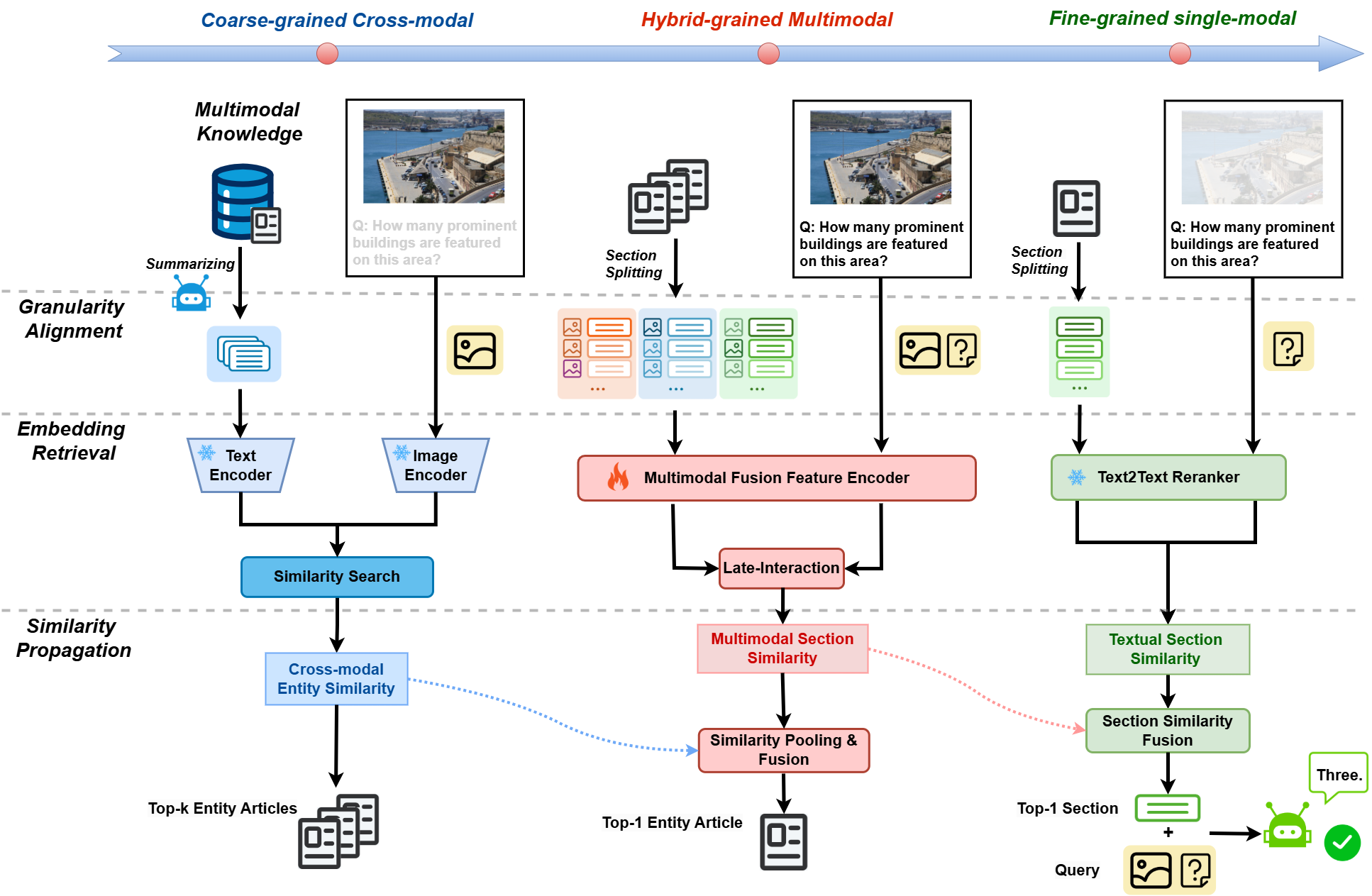}

    \caption{An overview of our framework. Our framework first performs a coarse-grained, cross-modal entity search by using offline-generated knowledge summaries as indices and a query image as the query; it then conducts a hybrid-grained multimodal-fused reranking that aligns image and text details via feature fusion and integrates cross-step similarities. Finally, a fine-grained section-augmented generation selects the most relevant knowledge section through combined text and multimodal similarities to support answer generation.}
    \label{fig:framework}
    \vspace{-4mm}
\end{figure*}

\section{Methodology}
% \section{Approach}

% We give an overview of our proposed multi-step framework.
% Given a reference image and knowledge-intensive question from user, our goal is to construct a visual question answering system

To address the challenges in KB-VQA, we introduce a multimodal RAG system that features efficient coarse-to-fine, multi-step multimodal retrieval. This system is capable of extracting pertinent information from a vast multimodal knowledge base with millions of entries. The retrieved data is subsequently used to enhance the generator's responses. An overview of the proposed framework is shown in Figure~\ref{fig:framework}, which comprises three key components: coarse-grained cross-modal entity searching, hybrid-grained multimodal-fused reranking, and fine-grained section-augmented generation.

\subsection{Coarse-Grained Cross-Modal Entity Searching}
To preliminarily filter out wiki entity information related to the query from a large corpus, we design a coarse-grained entity retrieval method. In this step, the query image indicating a subject serves as a coarse-grained query, while entity summaries act as coarse-grained candidates, ensuring appropriate information granularity alignment. 

\noindent \textbf{Summary Generation.}
Since the complete wiki entity articles are too redundant as entity information for effective retrieval indexing, we align them with the macro-level entity information of the query image. Specifically, we use article summaries as the retrieval index for wiki entities, these concise abstracts densely encapsulate key aspects of the entity's information. To maintain real-time retrieval efficiency, we generate summaries for all entity articles in the knowledge base offline. Given an entity article $a_i$, we employ a pre-trained language model $M_s$ with an instruction prompt template $P$ to generate a summary $s_i$, ensuring that the summary is both informative and well-aligned with the retrieval task. 

\begin{equation}
    s_i = M_s(P, a_i)
\end{equation}

\noindent \textbf{Image-to-Summary Entity Searching.}
After obtaining the wiki entity summaries, we use the query image from the VQA triplet, uniquely representing the target entity, as the query object for entity searching. Specifically, both the query image and candidate entity summaries are transformed into feature vectors, which are used for similarity-based matching for entity searching.
\begin{equation}
    Ent_k = F(E_v(I_q), E_t(S), k)
\end{equation}
where $I_q$ denotes the query image, and $S = {s_1, s_2, ..., s_n}$ represents all entity summaries in the knowledge base. These are encoded through the visual encoder $E_v$ and textual encoder $E_t$ of CLIP~\cite{radford2021learning}, respectively. We use the Faiss library $F$~\cite{johnson2019billion} to index the feature vectors of the entity summaries and perform embedding matching based on the inner product. Finally, we retain the top-$k$ most relevant entity summaries along with their corresponding entity information $Ent_k$.

\subsection{Hybrid-Grained Multimodal-Fused Reranking}

After the initial search, we obtain the top-k entity candidates most similar to the query subject. Within this candidate set, we extract hybrid-grained multimodal fusion feature matrices from the coarse-grained images and fine-grained texts provided by the query and candidate knowledge. Then, by employing a late-interaction mechanism, we obtain a fine-grained section similarity. By integrating the entity similarity from the previous step, we derive a coarse-grained reranking similarity that yields a more accurate ordering of relevant entities.

% we extract multimodal fusion feature matrices from both the candidate knowledge and the query's textual and visual information. These feature matrices are then used to compute the coarse-grained entity similarity and the fine-grained section similarity. The entities are reranked by integrating similarity scores from multimodal fusion-based retrieval step and the initial entity retrieval step.

\noindent \textbf{Multimodal Fusion Feature Matching.} Specifically, we leverage the Q-Former~\cite{li2023blip} architecture to extract multimodal fusion features from both query and candidates. The pipeline for fusion feature extraction is as follows:
\begin{equation}
    Q = E_m(q), C_{sec_e^h}= E_m(d_e^h)
\end{equation}
As the input to the q-former multimodal encoder $E_m$, we use the query image $I_q$ and textual question $T_q$ on the query side, forming the input pair $q = (I_q, T_q)$. On the candidate side, we construct the input as $d_e^h = (I_e, sec_e^h)$, where $I_e$ is the main image of entity and $sec_e^h$ is the $h$-th section of the entity article $a_e = \{ sec_e^1, sec_e^2, \dots, sec_e^p \}$. The fusion feature matrix $Q$ and $C_{sec_e^h}$ are the output vectors corresponding to the query tokens of encoder $E_m$.
% \begin{equation}
%     sim_m^{sec_e^h} = sim_m(q,d_e^h) = \sum_{i=1}^{l_Q} \max_{j=1}^{l_{C_{sec_e^h}}} Q_i {C_{sec_e^h}^j}^\top
% \end{equation}
We calculate the fine-grained multimodal similarity $sim_m^{sec_e^h}$ between the query $q$ and candidate $d_e^h$ using the late-interaction~\cite{khattab2020colbert}, which fully leverages the correlations between each token vector in the feature matrix and integrates them into a single similarity score through a Max-Sum operation. $l_Q$ and $l_C$ denote the total number of tokens in $Q$ and $C_{sec_e^h}$, respectively.
\begin{equation}
    sim_m^{sec_e^h} = sim_m(q,d_e^h) = \sum_{i=1}^{l_Q} \max_{j=1}^{l_C}Q_i {C_{sec_e^h}^j}^\top
\end{equation}
% \begin{equation}
%     e_1 = \arg\max_{e \in E_k} (\alpha \cdot sim_1^e + (1 - \alpha) \cdot \max_{h=1}^{l_{a_e}} sim_m^{sec_e^h})
% \end{equation}
Finally, we obtain the coarse-grained entity similarity by maximizing $sim_m$ of all the sections in the entity article $a_e$, with section size $l_{a_e}$. We then compute the final reranking similarity scores for the top-k entities $Ent_k$ by performing a weighted summation with the initial entity similarity $sim_c^e$ from the previous cross-modal retrieval step. The most relevant entity $e_{top1}$ is selected based on the highest reranking score. The weighting factor $\alpha$ balances the contributions of the two similarity measures, optimizing reranking performance.
% 1 entity similarity  to derive the reranking similarities scores for the top-k entities $E_k$, based on which we can retrieve the entity $e_1$ most relevant to the query. The $\alpha$ here is used to balance the two similarities in order to achieve the best reranking performance.
\begin{equation}
    e_{top1} = \arg\max_{e \in Ent_k} (\alpha \cdot sim_c^e + (1 - \alpha) \cdot \max_{h=1}^{l_{a_e}} sim_m^{sec_e^h})
\end{equation}

\noindent \textbf{Multimodal Reranker Training.}
To train the multimodal fusion encoder, we employ contrastive learning with the hard negative samples from previous retrieval step. Specifically, the coarse-grained retrieval in earlier step generates a top-k candidate entity set for each training sample. Most of these entities share similar coarse-grained characteristics, but differ in fine-grained details.
% For the multimodal-fused encoder used in this step, we train it using contrastive learning with the hard negative samples from step 1. Specifically, we use the coarse-grained retrieval of step 1 to generate a top-k candidate entity set for each training sample. Most of these entities are very similar at the coarse-grained knowledge but differ in fine-grained information. 
To construct training pairs, we randomly select negative pairs by pairing the main images of candidate entities with non-evidentiary sections from their articles. In contrast, positive pairs consist of the main image and the evidence section of the correct entity. 
% Next, we randomly select some pairs of main images and non-evidentiary sections from articles of the candidate set of each training sample as negative pairs, while the evidence section and the main image of its corresponding entity are used as the positive pair. 
\begin{equation}
    \mathcal{L} = - \log \frac{\exp(sim_m(q,d_{+}) / \mathcal{T})}{\sum_{j=1}^{N} \exp(sim_m(q,d_j) / \mathcal{T})}
\end{equation}
By constructing each training sample containing $N$ contrastive pairs $d$, we train the reranker to match queries with the positive candidate pairs $d_{+}$ based on hybrid-grained multimodal information with the adaptive temperature $\mathcal{T}$ for the smoothness of the softmax distribution. This multimodal-fused retrieval approach not only complements the coarse-grained entity retrieval in step one but also paves the way for subsequent fine-grained refinement.

\subsection{Fine-Grained Section-Augmented Generation}
After completing the first two steps of entity ranking, we identify $e_{top1}$ as the entity most relevant to the query. In this step, we perform fine-grained knowledge filtering on the entity information and use the filtered knowledge as auxiliary context to enhance the generation of the downstream generator. Specifically, we employ a pre-trained textual reranker $R_t$ to calculate the textual similarities $sim_t^{sec} = R_t(T_q, sec)$ between sections of the entity article and question. These textual
similarities are then combined with the fine-grained multimodal-fused similarity $sim_m^{sec}$ obtained from the previous step through weighted summation, enabling us to extract the most relevant entity section $sec_{e_{top1}}^{best}$ for the query.
\begin{equation}
    sec_{e_{top1}}^{best} = \arg\max_{sec \in e_{top1}} (\beta \cdot sim_m^{sec} + (1 - \beta) \cdot sim_t^{sec})
\end{equation}
Here, $\beta$ serves as a balancing factor to weigh the similarities derived from retrieval at different steps. Finally, the most relevant section is provided as input context along with the query to the generator for question-answering.
% , which helps mitigate hallucination and improves the accuracy of the responses.

 % The section-granularity text knowledge filtering primarily focuses on extracting information relevant to the textual query objectives. Therefore, pure text matching is used as the main filtering method. However, considering that multimodal combined features can provide information from different dimensions at a finer granularity, the fusion similarity approach is adopted.

% \section{Experimental Settings}
\section{Experiments}

\begin{table}[]
    \centering
    \resizebox{\columnwidth}{!}{%
    \small
    \begin{tabular}{l cccc c cc}
    \toprule
     \multirow{2}{*}{\textbf{Datasets}} & \multicolumn{4}{c}{\textbf{\#Samples}} & & \multicolumn{2}{c}{\textbf{\#Entity Articles}} \\
     \cmidrule{2-5} \cmidrule{7-8} 
    & Gen. Train & Ret. Train & Valid & Test & & Train & Valid/Test \\
    \midrule
    Infoseek   & 100K  & - & - & 71,335 & & 100K  & 100K     \\
    E-VQA      & 100K  & 191k & 11,696 & 4,750 & & 2M & 2M      \\
    \bottomrule
    \end{tabular}
    }
    \caption{Statistics of Infoseek and E-VQA datasets used in our experiments. Gen. Train and Ret. Train represent the number of training samples for the generator and the multimodal fusion module in reranking, respectively.
    }
    \label{tab:dataset_stats}
    \vspace{-4mm}
\end{table}

\subsection{Datasets}

We utilize two challenging KB-VQA datasets, InfoSeek and E-VQA~\cite{chen-etal-2023-pre-trained,mensink2023encyclopedic}, for training and testing. To ensure fairness in both retrieval and VQA evaluations, we adopt the same setup used by many previous studies, with the specific configuration detailed in Table~\ref{tab:dataset_stats}. For the images associated with the entity articles in the knowledge base, we crawled multiple images, including the main image, from each entity's corresponding wikipedia page to constitute the knowledge base’s image set. Our experiments primarily focus on assessing both retrieval and question-answer performance: retrieval is evaluated using Recall@K, while VQA performance is measured using the official metrics for each dataset (e.g., BEM score~\cite{zhang2019bertscore} for E-VQA and both VQA accuracy~\cite{antol2015vqa} and relaxed accuracy~\cite{methani2020plotqa} for InfoSeek). More details on the datasets and their evaluation methods can be found in the Appendix~\ref{sec:dataset details}.

 % Among these, the 100k samples for generator fine-tuning are randomly drawn from each dataset, while the fine-tuning samples for the multimodal fusion retriever are selected from all unique entity–question–answer triples in E-VQA, which have section annotations. The E-VQA's knowledge base consists of the original 2 million entities, whereas InfoSeek, similar to other works, employs 100k entities extracted from the original 6 million that contain all test entities.

\subsection{Implementation Details}
In this section, we briefly introduce some details of the various steps in our framework. Regarding the models and prompts used, please refer to Appendix~\ref{sec:model checkpoints} and Appendix~\ref{sec:prompts}, respectively.

\vspace{3pt} 
\noindent \textbf{Initial Entity Searching.}
To produce high-quality summaries of entity articles, we leveraged LangGPT~\cite{wang2024langgpt} to develop the "Wiki Summary Generator Assistant" prompt. For efficient entity search using the FAISS library, we employed the pooled embeddings from the last layer of encoder for calculating image-text similarity. We set $k$ to 20 and retrieved only the top-20 most relevant entities to strike a balance between the speed and effectiveness of subsequent retrieval and question-answering tasks, as validated by the experimental results presented in the Table~\ref{tab:ablation_retrieval_scope}.

\vspace{3pt} 
\noindent \textbf{Multimodal-Fused Encoder Training and Inference.}
The encoder model is initialized with pre-trained q-former weights using the LAVIS Library~\cite{li-etal-2023-lavis}. We select the top 32 embeddings from the model output as our multi-modal fusion feature matrix, corresponding to the position and number of defined query tokens. Since InfoSeek training samples do not include labels for the evidence section, our encoder is trained on the E-VQA training set and then tested on both datasets. This setting aligns with the training requirements and characteristics of the datasets, while also allows us to evaluate the model's generalization ability. Given that each question-answer pair in the training samples corresponds to multiple query images, we use only the first query image to form the triplet with the question-answer pair as the actual training sample. This strategy ensures training quality while enhancing efficiency. Additional details regarding this step can be found in Appendix~\ref{sec:step2 details}. Results of ablation experiments on the hyperparameter for cross-stage similarity propagation are presented in Appendix~\ref{sec:hyper-parameters' value}.

\vspace{3pt} 
\noindent \textbf{Generator Training and Inference.}
we primarily employ pre-trained LLM/MLLM models for ablation studies and most of the main experiments. To assess the performance of the generator fine-tuned under our retrieval system, we experiment with lightweight and efficient fine-tuning on LLaVA-1.5-7B~\cite{liu2024improved} and evaluate its VQA performance. More details are presented in Appendix~\ref{sec:step3 details}. 
% Ablation experiment results on the hyperparameter $\beta$ of cross-stage similarity propagation in Step 3 can be found in Appendix~\ref{sec:hyper-parameters' value}.

% \section{Experimental Results and Analyses}

\begin{table*}[t]
\fontsize{8}{10}\selectfont
  \centering
  \setlength{\tabcolsep}{3mm}
  \begin{tabular}{l cccc c cccc}
   \toprule
   %为最好的结果加粗
    \multirow{2}{*}{\textbf{Method}}& \multicolumn{4}{c}{\textbf{E-VQA}} & & \multicolumn{4}{c}{\textbf{InfoSeek}} \\
    \cmidrule{2-5} \cmidrule{7-10} 
     & R@1 & R@5 & R@10 & R@20 & & R@1 & R@5 & R@10 & R@20 \\
    \midrule
    CLIP I-T & 3.3 & 7.7 & 12.1 & 16.5 & & 32.0 & 54.0 & 61.6 & 68.2 \\
    %CLIP I-T indicates the retrieval from the reference image to Wikipedia entry texts with CLIP(result reference from Echosight)
    Wiki-LLaVA & 3.3 & - & 9.9 & 13.2 & & 36.9 & - & 66.1 & 71.9 \\
    LLM-RA & - & - & - & - & & 47.3 & 53.8 & - & - \\
    mR$^2$AG & - & - & - & - & & 38.0 & - & 65.0 & 71.0 \\
    ReflectiVA & 15.6 & 36.1 & - & \underline{49.8} & & \underline{56.1} & \underline{77.6} & - & \textbf{86.4} \\
    EchoSight \\
    \hspace{0.5em}\textit{w/o. reranking} & 13.3 & 31.3 & 41.0 & 48.8 & & 45.6 & 67.1 & 73.0 & 77.9 \\
    \hspace{0.5em}\textit{w. reranking} & \underline{36.5} & \underline{47.9} & 48.8 & 48.8 & & 53.2 & 74.0 & 77.4 & 77.9 \\ %
    OMGM (ours) \\
    \hspace{0.5em}\textit{w/o. reranking} & 19.1 & 41.2 & \underline{49.8} & 58.7 & & 52.6 & 73.9 & \underline{80.0} & 84.8 \\
    \hspace{0.5em}\textit{w. reranking} & \textbf{42.8} & \textbf{55.7} & \textbf{58.1} & \textbf{58.7} & & \textbf{64.0} & \textbf{80.8} & \textbf{83.6} & \underline{84.8} \\ % 
    %plus w rerank wo rerank
  \bottomrule
  \end{tabular}
  \caption{Retrieval results on the E-VQA test set and InfoSeek validation set. "w/o. reranking" and "w. reranking" represent the entity retrieval results after step one and step two, respectively. Best in bold, second-best underlined.}
  \label{tab:retrieval_main_results}
\end{table*}

%retrieval results
% E-VQA | InfoSeek
% Recall 1 5 10 20 | Recall 1 5 10 20
%Wiki-LLaVA 3.3 - 9.9 13.2 | 36.9 - 66.1 71.9
%mR^2AG - - - - | 38.0 - 65.0 71.0
%LLM-RA - - - - | - 47.31 - -
%EchoSight 36.5 47.9 48.8 48.8 | 53.2 74.0 77.4 77.9
%Ours 42.8 55.7 58.1 58.7 | 64.0 80.8 83.6 84.8

\begin{table*}[ht]
\fontsize{8}{10}\selectfont
\renewcommand{\arraystretch}{1.1}
\centering
\setlength{\tabcolsep}{1mm}
\begin{tabular}{l|c|c|c | c | ccc}
\toprule
%为最好的结果加粗
\multirow{2}{*}{\textbf{Method}} & \multirow{2}{*}{\textbf{Generator Model}} & \multirow{2}{*}{\textbf{Gen. FT}} & \multirow{2}{*}{\textbf{Ret. FT}} & \multirow{2}{*}{\textbf{E-VQA}} & \multicolumn{3}{c}{\textbf{InfoSeek}} \\
 & & & & & \textbf{Unseen-Q} & \textbf{Unseen-E} & \textbf{Overall} \\
\midrule
RoRA-VLM & LLaVA-1.5-7B & \Checkmark & \XSolidBrush & 20.29 & 27.34 & 25.10 & - \\ 
Wiki-LLaVA & LLaVA-1.5-7B & \Checkmark & \XSolidBrush & 21.8 & 30.1 & 27.8 & 28.9 \\
LLM-RA & BLIP2-Flan-T5XL & \Checkmark & \Checkmark & - & 26.12 & 20.90 & 23.14 \\
EchoSight & Mistral-7B | LLaMA3-8B & \XSolidBrush & \Checkmark & 41.8 & - & - & 31.3\\
mR$^2$AG & LLaVA-1.5-7B & \Checkmark & \Checkmark & - & \underline{40.6} & \underline{39.8} & \underline{40.2}\\
ReflectiVA & LLaVA-MORE-8B & \Checkmark & \Checkmark & 35.5 & 40.4 & 39.8 & 40.1\\

\midrule
\multirow{3}{*}{OMGM (ours)} & InternVL-2.5-8B  & \XSolidBrush & \Checkmark & 48.72 & 37.16 & 35.1 & 36.1 \\
& LLaMA3-8B & \XSolidBrush & \Checkmark & \underline{49.94} & 35.26 & 33.61 & 34.42\\ 
& LLaVA-1.5-7B & \Checkmark & \Checkmark & \textbf{50.17} & \textbf{43.46} & \textbf{43.53} & \textbf{43.49}\\ 

\bottomrule
\end{tabular}
\caption{VQA accuracy comparison with the baselines. Gen. FT and Ret. FT indicate whether the generator and retriever of the method were fine-tuned, respectively. Best in bold, second-best underlined.} 
\label{tab:vqa_main_results}
\vspace{-2mm}
\end{table*}

\subsection{Main Results}
The results of our method compared with other works are presented in Tables~\ref{tab:retrieval_main_results} and Table~\ref{tab:vqa_main_results}, primarily on entity retrieval and VQA performance.
% Additionally, we evaluate OMGM on OK-VQA, which lacks predefined document structure, to demonstrate its generalization. The detailed analyses is presented in the Appendix \ref{sec:ok-vqa ablation}.
% The detailed analyses are provided below.

\vspace{3pt} 
\noindent \textbf{Retrieval Result.} Table~\ref{tab:retrieval_main_results} showcases the retrieval performance of various multimodal RAG approaches across two datasets. The "CLIP I-T" refers to the naive approach, where CLIP is used for the cross-modal similarity search, linking the query image to the wiki article with embeddings. By examining the Recall@1 results, our method (w. reranking) significantly outperforms other methods, which demonstrates the robust retrieval capability of our retrieval method. Additionally, the reranking in step two improves Recall@1 by 23.7\% and 11.4\% for the two datasets compared to step one alone, underscoring the effectiveness of multimodal retrieval and cross-step similarity propagation in boosting reranking performance. Furthermore, even without reranking, our method outperforms the full-scale retrieval strategies commonly used in other works across all retrieval ranges, demonstrating the significant advantage provided by the granularity alignment between the summary and the query image in step 1.
% which indicates that the multimodal-used retrieval in step 2 and cross-step similarity propagation have effectively enhanced the reranking performance. 
% Moreover, the performance of our method (w/o. Reranking) outperforms the full-scale retrieval employed by most works across all retrieval ranges, demonstrating the positive impact of the granularity alignment between the summary and the reference image in Step 1 on retrieval.
%非重排优于其他方法,同粒度对齐的好
%重排带来很好的gain,多模态融合和跨步骤交互做的好

\vspace{3pt} 
\noindent \textbf{VQA Result.} Table~\ref{tab:vqa_main_results} presents a comparison of our VQA results with the state-of-the-art methods. RORA-VLM~\cite{qi2024rora} is a retrieval-augmented VLM system that removes irrelevant information based on token-level embedding similarity and incorporates noise-resilient retrieval-augmented training. mR$^2$AG~\cite{zhang2024mr} and ReflectiVA~\cite{cocchi2024augmenting} center on reflection. By fine-tuning large models, they leverage token outputs to drive the retrieval process and perform re-screening and modifications on retrieved content and generated answers. Our method achieves superior VQA results on both datasets compared to existing approaches, demonstrating the improved generation performance of downstream models after enhancing retrieval capabilities. Notably, our method, which only fine-tunes the retriever, outperforms most approaches that fine-tune downstream generators, highlighting its efficiency. We test the VQA performance of mainstream LLMs and MLLMs under zero-shot settings and our framework on two datasets and provide results in Appendix~\ref{sec:ablation_diff_gm}. Our method demonstrates excellent VQA results across both MLLM and LLM models, underscoring its ability to generalize retrieval optimization to enhance generation across different types of downstream generators.

%vqa acc 
%Finetune | E-VQA eval score | InfoSeek final unseen
%Ours(llama3): x | 49.94 | 34.42
%E-VQA echosight wikillava RORA-VLM as acc baseline（same single hop）
%InfoSeek wikillava echosight RORA-VLM as acc baseline

\subsection{Ablation Study}
We conduct extensive experiments to assess the effectiveness of our framework and the design of each step, focusing on retrieval and question-answering performance. Due to space limitations, we have placed additional ablation experiments in Appendix~\ref{sec:hyper-parameters' value} and Appendix~\ref{sec:ablation_step2_sim}, and the results of some ablation experiments on InfoSeek can be found in Appendix ~\ref{sec:infoseek ablation}. 
% We conducted many experiments to analyze the overall effectiveness of our framework and the rationality of each step's design through retrieval performance and downstream question answering results. The results of some ablation experiments on InfoSeek can be found in Appendix~\ref{sec:infoseek ablation}.

\vspace{3pt} 
\noindent \textbf{Impact of Retrieval Steps on VQA Results.} Table~\ref{tab:ablation_diff_steps} records the VQA results on two datasets, as each step of our framework is executed in sequence. When only step one is completed, the generator receives the top-1 entity article as context. After step two is executed, the generator's context is provided with the top-1 section.  Each step in our multi-step framework progressively improves the VQA performance of the downstream generator, confirming the effectiveness of its design.
Notably, step two offers the most significant enhancement, highlighting the powerful retrieval capability of our multimodal-fused reranker with similarity propagation.
% Overall, each step of the multi-step framework enhances the VQA performance of the downstream generator, validating the rationality of its design. 
% By observing the extent of improvement at each step, we can see that step 2 provides the greatest enhancement, which also indicates the powerful retrieval capability of multimodal-fused retrieval with similarities propagation in our approach.
\begin{table}[ht]
\small
\setlength{\tabcolsep}{2.5mm}
\fontsize{8}{10}\selectfont
\centering
\begin{tabular}{c c c | c c}
\toprule
step-1 & step-2 & step-3 & E-VQA & InfoSeek \\
\midrule
\XSolidBrush & \XSolidBrush & \XSolidBrush & 17.2 & 9.3 \\
\Checkmark & \XSolidBrush & \XSolidBrush & 25.45 & 18.87 \\% 8.25 9.57
\Checkmark & \Checkmark & \XSolidBrush & 39.81 & 31.38 \\ % 14.36 12.51
\Checkmark & \Checkmark & \Checkmark & 41.81 & 33.29 \\% 2 1.91
%plus 每个阶段上升的值
\bottomrule
\end{tabular}
\caption{The ablation study on the impact of different steps in our framework on the VQA results, using LLaVA-1.5-7B as the answer generator.}
\label{tab:ablation_diff_steps}
\vspace{-4mm}
\end{table}

\vspace{3pt} 
\noindent \textbf{Impact of Retrieval Design on Initial Entity Search. } As the first step of multi-step retrieval, we perform a coarse-grained search for entity-related knowledge within a large knowledge base. We use the query image as the query, which uniquely contains entity information. Retrieval candidates can be selected based on different modalities and knowledge granularities. Table~\ref{tab:ablation_step1_evqa} presents our retrieval experiments using images, articles, titles, and summaries as candidates. The results indicate that the Image-to-Summary method achieves the best retrieval performance, suggesting that summaries offer better alignment with query images in terms of information granularity, leading to improved retrieval outcomes.
% Based on the full-range recall results, it can be concluded that the Image-to-Summary method used in step 1 of our framework achieves the best retrieval performance. This indicates that summaries are more aligned with reference images in terms of information granularity than other forms of entity knowledge, thereby leading to superior retrieval performance.

\begin{table}[ht]
\small
\setlength{\tabcolsep}{2mm}
\fontsize{8}{10}\selectfont
\centering
\begin{tabular}{c | c c c c}
\toprule
Ret. Method & R@1 & R@5 & R@10 & R@20 \\
\midrule
$Image \rightarrow Article$ & 13.2 & 27.7 & 35.5 & 41.7\\
$Image \rightarrow Image$ & 13.4 & 31.8 & 41.9 & 48.8\\
$Image \rightarrow Title$ & 17.5 & 31.9 & 38.6 & 44.8\\
\textbf{$Image \rightarrow Summary$} & \textbf{19.1} & \textbf{41.2} & \textbf{49.8} & \textbf{58.7}\\
\bottomrule
\end{tabular}
\caption{Ablation study on modalities and granularities design for entity retrieval in step 1.}
\label{tab:ablation_step1_evqa}
\end{table}

\vspace{3pt} 
\noindent \textbf{Effect of Multimodal Fusion Reranking.} In step two, we rerank a small subset of entities obtained from the initial search to identify the most relevant one to the query. Consequently, we must leverage question-oriented, fine-grained textual information, such as the query question and the sections associated with each entity. In VQA tasks, both the query and the knowledge base are inherently multi-modal. Therefore, it is reasonable to consider incorporating multi-modal fusion features into the retrieval process. Based on this rationale, we experimented with four retrieval approaches: purely text-based retrieval, multimodal-to-text retrieval, text-to-multimodal retrieval, and fully multimodal retrieval. Using Q-Former as the base encoder model, we fine-tune all four retrieval modality approaches under the same training configuration and evaluated their performance. The results presented in Table~\ref{tab:ablation_step2_modality_evqa} indicate that the method employing multimodal fusion features on both the query and candidate sides achieve the best retrieval performance. This finding directly demonstrates the effectiveness and comprehensiveness of the multimodal-fused reranking design in our framework.

\begin{table}[ht]
\small
\setlength{\tabcolsep}{2mm}
\fontsize{8}{10}\selectfont
\centering
\begin{tabular}{c | c c c c}
\toprule
Ret. Modality & Sec. R@1 & R@1 & R@5 & R@10 \\
\midrule
$T \rightarrow T$ & 24.6 & 30.7 & 51.8 & 57.4 \\% Sec/R@1=80%
$(I,T) \rightarrow T$ & 22.5  & 28.7 & 51.3 & 57.0  \\% 78%
$T \rightarrow (I,T)$ & 24.3 & 30.3 & 51.1 & 56.9 \\% 80%
\textbf{$(I,T) \rightarrow (I,T)$} & \textbf{32.8} & \textbf{40.2} & \textbf{54.8} & \textbf{57.8} \\% 81.6%
\bottomrule
\end{tabular}
\caption{Ablation study on retrieval modality of step two on E-VQA. "I" indicates image modality and "T" indicates text modality. "Sec. R@1" refers to the recall of the top-1 section. $\alpha$ is set to 0 for direct comparison.}
\label{tab:ablation_step2_modality_evqa}
\vspace{-4mm}
\end{table}

\vspace{3pt} 
\noindent \textbf{Effect of Reranking Scope on Retrieval Performance.} The reranking scope $k$ influences both the number of entities filtered during the initial search and the range of multi-modal fused features extraction during reranking. Table~\ref{tab:ablation_retrieval_scope} presents the retrieval results and the average retrieval time for the first two steps with varying $k$ values. Our framework can consistently improve the retrieval capacity while increasing k from 10 to 100, accompanied with more retrieval time as well. To balance retrieval quality and efficiency, we set the reranking scope $k$ to 20 for experiments across both datasets.
% And in consideration of balancing retrieval time and quality, we set the reranking scope k to 20 when conducting experiments on both datasets.

\begin{table}[ht]
\small
\setlength{\tabcolsep}{1mm}
\centering
\begin{tabular}{c | cccccc c}
\toprule
k & Sec. R@1 & R@1 & R@5 & R@10 & R@20 & Time \\
\midrule
10 & 31.6 & 39.0 & 48.8 & 49.8 & - & 0.630\\
20 & 34.7 & 42.8 & 55.7 & 58.1 & 58.7 & 1.110\\
50 & 37.4 & 45.9 & 61.3 & 64.6 & 66.7 & 2.420\\
100 & 38.8 & 47.4 & 65.0 & 69.4 & 72.2 & 4.642\\

\bottomrule
\end{tabular}
\caption{The effect of the retrieval scope K on the retrieval results and time after step two on E-VQA.}
\label{tab:ablation_retrieval_scope}
\vspace{-4mm}
\end{table}

\noindent \textbf{Generalization on Additional Benchmark with Differen Document Structures.} For E-VQA and InfoSeek documents, which consist of entity-based Wikipedia articles, we segmented them into sections based on their inherent structure. This method allowed us to leverage the natural organization of Wiki-style content, which often includes headings and subheadings, to create meaningful sections for retrieval. For KB-VQA datasets with document structures that differ significantly from E-VQA or InfoSeek, alternative segmentation strategies can be employed to ensure effective organization of knowledge. One potential approach is to use rule-based method, which relies on predefined heuristics such as paragraph breaks, headings, or specific keywords to define section boundaries. Another approach is semantic clustering, which groups text segments based on semantic similarity, enabling the creation of fine-grained, section-like knowledge units even in the absence of explicit document structure. 

To evaluate the generalizability of OMGM on different segmenation setting, we tested our method on the OK-VQA\cite{marino2019ok} dataset, which lacks a highly structured format and segmented using a rule-based segmentation method. As shown in Table \ref{tab:okvqa_eval}, OMGM demonstrated strong performance, achieving higher retrieval accuracy (Pseudo Recall@5) and VQA scores compared to the strong baseline PreFLMR\cite{lin-etal-2024-preflmr}. Specifically, OMGM achieved a Pseudo Recall@5 of 73.4 and a VQA score of 66.57, outperforming PreFLMR by a notable margin. These results suggest that OMGM’s framework is robust and adaptable to varying document structures across different KB-VQA datasets.

\begin{table}[ht]
\small
\setlength{\tabcolsep}{2mm}
\fontsize{8}{10}\selectfont
\centering
\begin{tabular}{c | c c }
\toprule
Method & Pseudo Recall@5 & VQA score \\
\midrule
PreFLMR & 70.9 & 61.88 \\
\textbf{OMGM} & \textbf{73.4}  & \textbf{66.57}  \\
\bottomrule
\end{tabular}
\caption{Comparison of the retrieval and VQA results of OMGM and PreFLMR on OK-VQA.}
\label{tab:okvqa_eval}
\end{table}

\noindent \textbf{Quantitative evaluation on the efficiency of OMGM's step-by-step approach.} To further perform the efficiency of the proposed OMGM framework, we conducted a comparative evaluation of OMGM, the one-step multimodal RAG method PreFLMR\cite{lin-etal-2024-preflmr}, and the direct use of LLaVA-1.5-7B for VQA on the E-VQA dataset. We evaluated three key metrics: average retrieval time, average inference time, and VQA performance. It is worth noting that PreFLMR preprocesses and encapsulates all passage embeddings prior to inference, which reduces retrieval time during runtime to primarily consist of query embedding and similarity matching. To ensure a fair comparison, OMGM’s retrieval time calculation also focuses on these two components.

\begin{table}[ht]
\small
\setlength{\tabcolsep}{1.5mm}
\fontsize{8}{10}\selectfont
\centering
\begin{tabular}{c | c c c}
\toprule
Method & Avg. Ret. Time & Avg. Inf. Time & VQA Result \\
\midrule
LLaVA-1.5-7B & - & 1.432 & 17.00 \\
PreFLMR & 0.984 & 2.196 & 54.45 \\
\textbf{OMGM} & \textbf{0.402}  & \textbf{2.023} & \textbf{63.39} \\
\bottomrule
\end{tabular}
\caption{Comparison of average retrieval and inference time as well as VQA performance on E-VQA.}
\label{tab:ablation_stepbystep_latency}
\end{table}

As shown in Table \ref{tab:ablation_stepbystep_latency}, OMGM achieves substantial improvements in VQA performance, surpassing both LLaVA-1.5-7B and PreFLMR. Despite employing a step-by-step retrieval strategy, OMGM maintains competitive inference efficiency. Compared to direct use of LLaVA-1.5-7B, OMGM delivers significantly better VQA results while introducing only a minimal increase in inference time. 

Additionally, when compared to the one-step PreFLMR, OMGM demonstrates a notable reduction in retrieval time, decreasing from 0.98s to 0.4s. This improvement is due to its orchestrated retrieval process, which is specifically designed to optimize the integration of different modalities and knowledge granularities at each step, achieving an effective balance between retrieval efficiency and performance.

\section{Conclusion}
\vspace{3pt} 
In this paper, we propose a RAG system with multi-step multimodal retrieval. By employing queries and candidates of appropriate modalities at each step, the system aligns the information granularities for better retrieval. Our system capitalizes on cross-step similarity propagation to enhance retrieval interactions and employs a multimodal-fused design to fully exploit the rich multimodal information present in queries and candidates. Experimental results on mainstream KB-VQA datasets show that our approach surpasses existing approaches in retrieval performance. The comprehensive retrieval pipeline enables pre-trained models and lightly fine-tuned models to outperform the systems heavily reliant on extensive fine-tuning, and we reveal the rationality and effectiveness of it by ablation studies. These findings provide valuable insights for designing effective multimodal retrieval systems tailored to KB-VQA tasks.

\section*{Limitations}
Our RAG system focuses primarily on the design of the multimodal retrieval module. While it has demonstrated significant performance improvements, several limitations remain:
1) Multimodal knowledge bases often include not only coarse-grained main images of entities but also numerous fine-grained secondary images linked to specific sections. Our current method does not exploit these secondary images to optimize multimodal retrieval, presenting a promising area for future exploration. 
2) Although our approach enhances VQA performance through improved retrieval, further exploration is needed to determine how generator models can more effectively utilize multimodal fusion features to enhance answer quality during the generation phase.
% The current framework is limited by the lack of suitable KB-VQA datasets for effectively studying coarse-grained multimodal retrieval. Many existing datasets (e.g., E-VQA, InfoSeek) rely heavily on template-generated questions with low informational diversity, reducing their utility for training robust embedding models. Additionally, they fail to capture real-world scenarios where textual queries provide critical complementary information to images. This gap hinders the development of retrieval methods that can adapt to varying query informativeness and domain specificity. Future work will focus on exploring more diverse and realistic datasets to enable dynamic, adaptive retrieval strategies better aligned with real-world multimodal reasoning tasks.

% \section*{Acknowledgments}

% We thank the anonymous reviewers for their feedback on this work. This work was supported by ... .

% In addition to the coarse-grained main images of entities, multimodal knowledge bases usually contain many fine-grained secondary images, which are generally associated with certain fine-grained sections. Our method did not utilize this part to optimize the multimodal retrieval, which is a potential area for future exploration. 2) Although our method achieves superior VQA performance through optimized retrieval processes, there remains potential for exploring how generator models can leverage multimodal fusion features to enhance visual question answering during the answer generation phase.

\bibliography{anthology,custom}
%\bibliography{custom}

\clearpage
\appendix

\section*{Appendix}

\section{Prompt Used in Our Methodology}
\label{sec:prompts}

In this section, we present the prompt templates utilized for invoking LLM/MLLM in our methodology, encompassing processes such as LLM-based summary generation and VQA across various datasets. The prompt template used to instruct the LLM to generate summaries offline for entity knowledge is presented in Table~\ref{tab:summary_generator_prompt}. It is evident that the entire template is structured in an agent-like format. Under this format, we expect the LLM to strictly adhere to the instructions to efficiently produce high-quality entity summaries.

During the question-answering phase in step 3, we have designed corresponding prompt templates tailored to the characteristics of different datasets and the capabilities of LLM/MLLM. This is to ensure that the true efficacy of our method in question-answering is demonstrated as accurately as possible. As demonstrated by the prompt for E-VQA testing shown in the Table~\ref{tab:evqa_vqa_prompt}, when deploying the MLLM, simple visual constraints and a query image for the query are incorporated into the prompt. For the InfoSeek dataset, which employs the stringent exact match evaluation criteria, it is crucial to thoroughly guide the downstream generator to adhere to the prescribed output format in order to showcase genuine question-answering performance. Consequently, we have incorporated additional format instructions and a one-shot example from the training set into the prompt presented in Table~\ref{tab:infoseek_vqa_prompt}.

\section{Dataset Details}
\label{sec:dataset details}

\noindent \textbf{Encyclopedic VQA~\cite{mensink2023encyclopedic}} 
The dataset encompasses approximately 221k question-answer pairs linked to 16.7k distinct fine-grained entities, with each entity represented by up to five images. The fine-grained entities and associated images are derived from the iNaturalist 2021 dataset and the Google Landmarks Dataset V2~\cite{van2021benchmarking, weyand2020google}. Moreover, the dataset provides a controlled knowledge base derived from WikiWeb2M~\cite{burns-etal-2023-suite} with 2M Wikipedia articles with images, which contain evidences to support each answer. According to the number of reasoning steps required, the questions in the dataset can be divided into single-hop and two-hop questions.  The dataset triplets are split into training, validation, and test subsets, containing 1 million, 13k, and 5,800 samples, respectively. For performance comparison with other related works, we also only use single-hop questions for training and testing. Therefore, we adopt the sample allocation method as shown in the Table~\ref{tab:dataset_stats}.

To evaluate the performance of our proposed retrieval-augmented QA LLM framwork in E-VQA, we utilize the standard metric Recall@K and BEM score~\cite{zhang2019bertscore} as metrics to evaluate its retrieval capability and question-answering capability, respectively. Recall@K evaluates the proportion of test samples whose top-k retrieved entities contain the correct entity, thereby reflecting the retrieval performance within the top-k scope. As the specific evaluation metric for the E-VQA dataset, the BEM score is obtained by comparing the predicted answer with the correct answers using a BERT model specifically fine-tuned for answer similarity assessment. This method correctly evaluates candidate answers that are valid but do not exactly match the reference answers in annotations, as opposed to common VQA metrics.

\vspace{3pt} 
\noindent \textbf{InfoSeek~\cite{chen-etal-2023-pre-trained}}
The dataset comprises 1.3 million image-question-answer triplets, corresponding to approximately 11,000 visual entities from OVEN~\cite{hu2023open}. There are 8.9K human-written visual info-seeking questions and 1.3M automated generated questions in InfoSeek. The triplets are partitioned into training, validation, and test sets, containing approximately 934k, 73k, and 348k samples, respectively. Due to the lack of ground truth for test split, our evaluation is conducted on the validation set. In particular, both the validation and test sets feature questions pertaining to unseen entities or queries that are not encountered during training. Additionally, the dataset includes a knowledge base consisting of 6 million Wikipedia entities. To be consistent with related works, like EchoSight~\cite{yan-xie-2024-echosight}, we utilize the subset of 100,000 entities, ensuring the inclusion of the 6,741 entities corresponding to the questions from the training and validation splits. When collecting images from Wikipedia pages for Wikipedia entities, we found that a very small portion of validation samples in the infoseek dataset corresponded to correct entities that lacked associated images. As a result, we filter out these samples and conduct evaluations on the remaining 71,335 validation samples, which still account for 96.9\% of the original dataset and ensures that the final results are not significantly affected.

For the retrieval evalution, we retained Recall@K as the evaluation metric, consistent with E-VQA. Following the specific question-answering evaluation criteria of InfoSeek, we employed two different metrics based on the question types. For questions requiring string-based answers, such as entity names, we report accuracy using the VQA accuracy metric~\cite{antol2015vqa}. This metric allows for multiple valid answers by considering slight variations in phrasing as correct. The model is evaluated based on whether its answer exactly matches any of these valid responses. For questions requiring numeric answers, we use relaxed accuracy~\cite{methani2020plotqa}, which considers an answer correct if it falls within an acceptable tolerance range around the ground truth.

\section{More Experimental Details}
\label{sec:experimental details}

\subsection{Model Checkpoints.}
\label{sec:model checkpoints}
We adopted LLaMA-8B-Instruct~\cite{llama3modelcard} as the summarization model, whose robust ability to follow instructions guarantees the reliability of the output, while its open-source nature and compact size ensure high efficiency for a massive knowledge base. With the Eva-CLIP encoder (Eva-CLIP-8B)~\cite{sun2024eva}, we extract the embeddings of the query images and entity summaries. When selecting the best section from the entity article in step 3, we used BGE-Reranker-v2-m3~\cite{chen-etal-2024-m3} for obtaining text similarity, which is a lightweight, efficient text reranking model that has been pre-trained for optimal performance. Regarding the answer generator, we tested not only LLMs such as LLaMA-8B-Instruct and Mistral-7B-Instruct-v0.2, but also MLLMs like LLaVA-1.5-7B and InternVL-2.5-8B~\cite{llama3modelcard, jiang2023mistral, liu2024improved, chen2024internvl}. Additionally, we further tested the performance of GPT-4 and GPT-4-o, which refer to GPT-4-1106 and GPT-4-o-2024-08-06, respectively.

\subsection{More Details about Step 2 Experiments.}
\label{sec:step2 details}
When employing hard negative sampling for multimodal-fused retriever training, we set $N$ = 16 image-section pairs for one training sample, which contain only one positive pair and 15 negtive pairs. For the image in the positive pair, we offline select the most similar image to the query image from multiple wiki images associated with the evident entity, using Eva-CLP for similarity ranking. And we directly choose the first image of the entity article as the image for the negative pair. Among the 15 negative pairs, we select up to three harder negative pairs from other sections of the article that contains the evidence section. Through the above contrastive pairs construction method, we efficiently obtain positive pairs most relevant to the query image and question, as well as diverse and challenging negative pairs. During retrieval training, we use learning rate 1e-5 and batch size 8, training for 1 epoch on a total of 191k training samples processed from E-VQA. This configuration allow the training to complete in 11 hours on 1 Nvidia A100 (80G). In the reranking inference, we set the similarity fusion hyper-parameter $\alpha$ to 0.9, indicating that fine-grained multimodal similarity can be used to fine-tune the coarse-grained entity similarity, which can be valiated by the results as shown in Appendix~\ref{sec:hyper-parameters' value}

\subsection{More Details about Step 3 Experiments.}
\label{sec:step3 details}
For the best section similarity fusion, we set $\beta$ to 0.2 to incorporate multimodal information with a small weight into unimodal information, which can be proved by the results as presented in Appendix~\ref{sec:hyper-parameters' value}. For the LLaVA-1.5-7B fine-tuning samples, we randomly selected 100k samples from the training set of each dataset and used our system's retrieval step to match the corresponding section paragraphs, thereby constructing training samples in the RAG format. Regarding the fine-tuning settings, we adopted the official LLaVA-1.5 LoRA training parameters, using a learning rate of 2e-5 and a batch size of 8x16. Lightweight training was performed for 1 epoch on 1 Nvidia A100 (80G).

Regarding the baseline details in our main results, we consider mR²AG and ReflectiVA to have fine-tuned their MLLM to perform a reranking-like relevance reflection on the retrieved content, and thus we treat them as having fine-tuned the retriever. Additionally, the E-VQA results for mR²AG were not adopted because they utilize the online Google Lens retrieval results provided by E-VQA, which differ from the settings in other works where the retrieval system is custom-designed, so no comparison is made.

%model checkpoints

%hyperparameter in training and inference

\section{More Experiments and Ablation studys}
\label{sec:more_exps}

\subsection{Consistency of our work across LLMs and MLLMs.} 
\label{sec:ablation_diff_gm}
As shown in Table~\ref{tab:ablation_diff_gm}, we tested the VQA performance of mainstream LLMs and MLLMs under zero-shot settings and our framework on two datasets. Firstly, from the perspective of closed-source and open-source models, we can observe that our framework enhances the performance of small open-source models to approach or even surpass that of powerful closed-source models in a zero-shot setting (some cases are shown in Appendix~\ref{sec:case study}), while also significantly improving the VQA performance of closed-source models, which demonstrates the generalizability of our RAG system for generation models. Additionally, we observed that for LLMs, if the retrieval performance is sufficiently good, their question answering results can approach or even surpass those of MLLMs (49.94 for LLaMA3-8B and 48.72 for  InternVL-2.5-8B in E-VQA), when query images are not accessible. This highlights the correct retrieved content about visual entity can help  LLMs in analyzing and answering visual questions.

\begin{table}[t]
\small
\setlength{\tabcolsep}{3mm}
\fontsize{8}{10}\selectfont
\centering
\begin{tabular}{l | c | c c}
\toprule
Generator Model & W/OMGM & E-VQA & InfoSeek \\
%plus 使用我们方法提高的值
\midrule
\multirow{2}{*}{LLaVA-1.5-7B} & \XSolidBrush & 17.2 & 9.3 \\
& \Checkmark & 41.81 & 33.29 \\
\midrule
\multirow{2}{*}{InternVL-2.5-8B } & \XSolidBrush & 25.56 & 10.95 \\
& \Checkmark & 48.72 & 36.1 \\
\midrule
\multirow{2}{*}{Mistral-7B } & \XSolidBrush & 17.71 & 0 \\
& \Checkmark & 48.08 & 17.14 \\
\midrule
\multirow{2}{*}{LLaMA3-8B } & \XSolidBrush & 18.78 & 1.56 \\
& \Checkmark & 49.94 & 34.42 \\
\midrule
\multirow{2}{*}{GPT-4 } & \XSolidBrush & 18.23 & - \\
& \Checkmark & 50.88 & - \\
\midrule
\multirow{2}{*}{GPT4-o } & \XSolidBrush & 35.92 & 38.35 \\
& \Checkmark & 51.18 & 42.09 \\
\bottomrule
\end{tabular}
\caption{The ablation study on the impact of our method on the VQA results of different generator models. We use the Overall Score as the primary metric for the VQA results on the InfoSeek dataset. Regarding the performance variations of closed‐source models on InfoSeek, due to the vast number of samples in the InfoSeek test set and the cost constraints of API calls, we only evaluated the most critical model, GPT4-o.}
\label{tab:ablation_diff_gm}
\end{table}

\subsection{Effect of different hyper-parameters' value in step 2 and 3}
\label{sec:hyper-parameters' value}

In this section, we investigate the impact of the hyperparameters $\alpha$ and $\beta$, which control the similarity mixing in step 2 and 3 of our method, on the retrieval and question-answering results.

\begin{figure}[th]
    \centering
    \includegraphics[width=0.95\columnwidth]{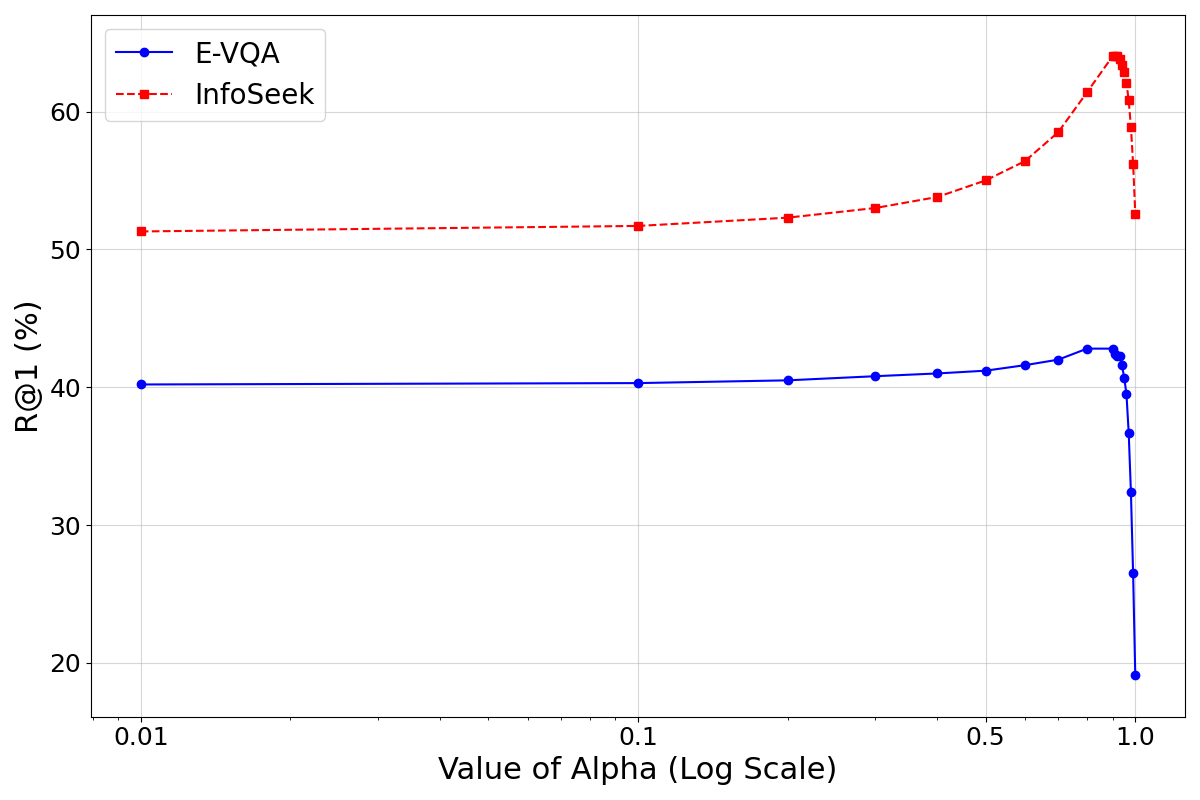}
    \caption{The variation of Entity Recall@1  with the change of step 2 similarity fusion hyper-parameter $\alpha$ on the E-VQA and InfoSeek }
    \label{fig:alpha_log}
\end{figure}

In step 2, we set $\alpha$ as the weight to integrate the coarse-grained similarities obtained from the initial entity search into the multimodal-fused similarities, thereby achieving a comprehensive reranking similarity. In the experiments with $\alpha$, we tested its variation from 0 to 1 and observed the changes in the final retrieval entity recall of step 2, as illustrated in the Figure ~\ref{fig:alpha_log}.

From the Figure ~\ref{fig:alpha_log}, we can see that the fusion of similarities at different steps has a very positive impact on the reranking effect. When $\alpha$ is set to 0, we solely rely on the similarity from step 2 for reranking, achieving relatively satisfactory results on both datasets (51.3\% and 40.2\%), highlighting the excellent entity reranking capability of the multimodal-fused reranker. Notably, for E-VQA, the retrieval outcomes in step 2 significantly surpass those of step 1, which is attributed to the fact that this dataset was used as the training set for the reranking encoder, demonstrating the remarkable effectiveness of multimodal fusion retrieval. Moreover, on InfoSeek, the approach exhibits strong generalizability, as combining similarities results in a substantial improvement in reranking performance (52.6\% to 64.0\%). The best retrieval performance is achieved when the similarity from step 1 is mixed at a higher proportion (0.9), indicating that the similarity of multimodal-fused retrieval optimizes the coarse-grained entity similarity obtained from the initial search in a "fine-tuning" manner.

\begin{figure}[th]
    \centering
    \noindent\includegraphics[width=0.95\columnwidth]{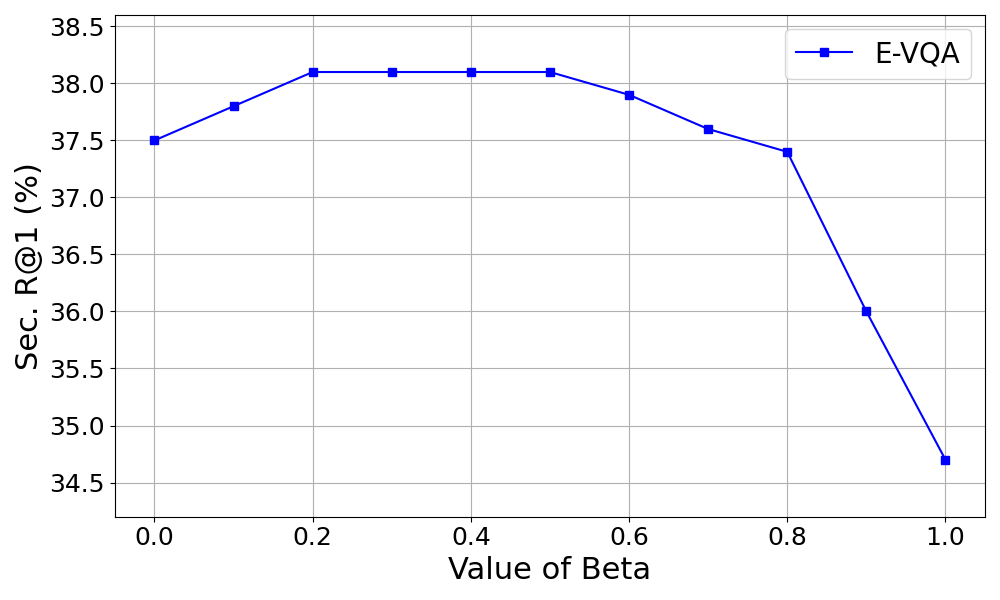}
    \caption{The changes in Section Recall@1 on the E-VQA dataset under varying values of the step 3 similarity fusion hyper-parameter $\beta$ }
    \label{fig:beta_evqa}
\end{figure}

\begin{figure}[th]
    \centering
    \noindent\includegraphics[width=0.95\columnwidth]{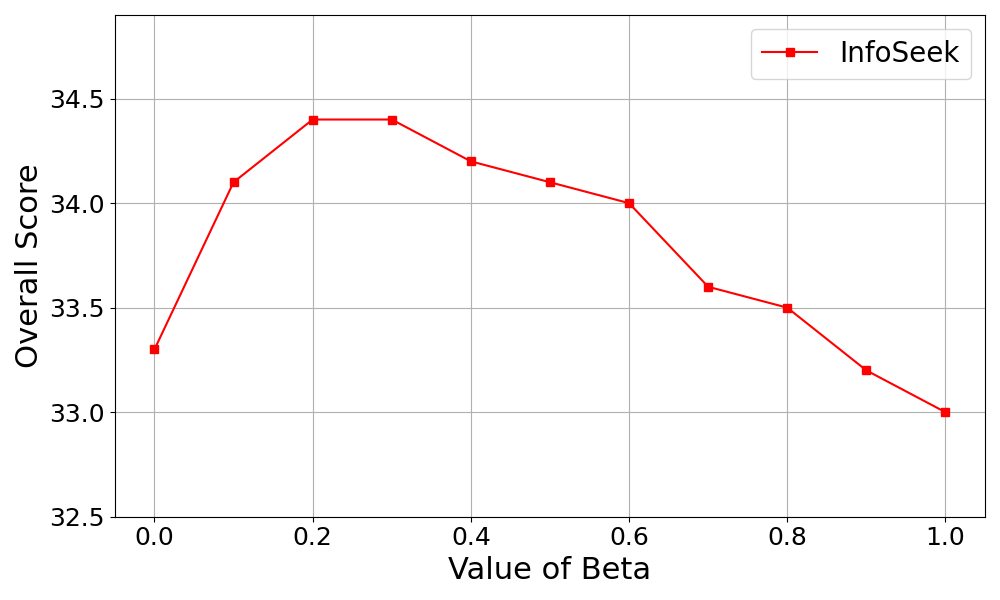}
    \caption{The changes in Overall Score on the InfoSeek dataset under varying values of the step 3 similarity fusion hyper-parameter $\beta$ }
    \label{fig:beta_is}
\end{figure}

In the step 3, we set $\beta$ as the weight to integrate the multimodal-fused section similarity into the direct text similarity, aiming for better knowledge denoising. In the experiments with $\beta$, we tested its variation from 0 to 1 and observed the changes in the retrieval and question-answering effectiveness on two datasets in step 2, as presented in Figure ~\ref{fig:beta_evqa} and Figure ~\ref{fig:beta_is}.

Compared to the significant impact of $\alpha$ on retrieval, the positive impact of $\beta$ is relatively limited. This is because the knowledge denoising in step 3 is conducted in a top-1 document context, without involving entity-level filtering, so simply using a pretrained text reranker can achieve good denoising results. The performance improvement obtained by integrating the multimodal fusion similarity at a small proportion also demonstrates the comprehensiveness and generalization ability of cross-step similarity propagation.

\subsection{Methods for computing multimodal fusion feature similarities}
\label{sec:ablation_step2_sim}
In order to achieve the best retrieval performance, we need to adopt an appropriate similarity computation method to match the multimodal fusion feature matrix. Based on previous studies~\cite{yan-xie-2024-echosight, radford2021learning, khattab2020colbert}, we evaluated three commonly used methods for computing similarity in Table~\ref{tab:ablation_step2_sim_evqa} and Table~\ref{tab:ablation_step2_sim_is}. Q-Former’s Image-to-Text Correspondence (ITC) computes the highest pairwise similarity between each multimodal query token embedding and the pooling token embedding of the multimodal candidate. CLIP’s ITC computes the similarity between the first token embeddings of the multimodal query and candidate. Late-Interaction gets the retrieval score by aggregating the maximum dot products over all query tokens with respect to all candidate tokens. From the results in the Table~\ref{tab:ablation_step2_sim_evqa} and Table~\ref{tab:ablation_step2_sim_is}, it is clear that, whether at the entity level or the section level, retrieval performance obtained through Late-Interaction is superior. This finding indicates that a fine-grained, comprehensive token-level similarity computation method is more suitable for  multimodal-fused retrieval.

\begin{table}[t]
\small
\setlength{\tabcolsep}{2mm}
\fontsize{8}{10}\selectfont
\centering
\begin{tabular}{c | c c c c}
\toprule
Sim. Calculation & Sec. R@1 & R@1 & R@5 & R@10 \\
\midrule
CLIP’s ITC & 28.3 & 37.1 & 53.2 & 57.5 \\
Q-Former’s ITC & 25.2 & 34.3 & 52.6 & 57.2 \\
Late-Interaction & \textbf{32.8}  & \textbf{40.2} & \textbf{54.8} & \textbf{57.8} \\
\bottomrule
\end{tabular}
\caption{The ablation study on the impact of different Step 2 similarity computation methods on entity-level and section-level retrieval results of E-VQA. The $\alpha$ of step 2 is set to 0 to facilitate a direct comparison.}
\label{tab:ablation_step2_sim_evqa}
\end{table}

\begin{table}[t]
\small
\setlength{\tabcolsep}{3mm}
\centering
\begin{tabular}{c | c c c}
\toprule
Sim. Calculation & R@1 & R@5 & R@10 \\
\midrule
CLIP’s ITC & 47.6 & 76.2 & 82.2 \\
Q-Former’s ITC & 47.5 & 76.6 & 82.4 \\
Late-Interaction & \textbf{51.3} & \textbf{77.9} & \textbf{82.8} \\
\bottomrule
\end{tabular}
\caption{The ablation study on the impact of different step 2 similarity computation methods on entity-level retrieval results of InfoSeek. The $\alpha$ of step 2 is set to 0 to facilitate a direct comparison.}
\label{tab:ablation_step2_sim_is}
\end{table}

\subsection{Some Ablation study results in InfoSeek} 
\label{sec:infoseek ablation}

\begin{table}[t]
\small
\setlength{\tabcolsep}{2mm}
\centering
\begin{tabular}{c | c c c c}
\toprule
Ret. Method & R@1 & R@5 & R@10 & R@20 \\
\midrule
$Image \rightarrow Article$ & 44.5 & 64.5 & 70.7 & 76.0\\
$Image \rightarrow Image$ & 45.6 & 68.6 & 74.6 & 77.9\\
$Image \rightarrow Title$ & 51.5 & 69.2 & 74.8 & 79.1\\
\textbf{$Image \rightarrow Summary$} & \textbf{52.6} & \textbf{73.9} & \textbf{80.0} & \textbf{84.8}\\
\bottomrule
\end{tabular}
\caption{The ablation study on the impact of different step 1 retrieval methods on entity retrieval results of InfoSeek.}
\label{tab:ablation_step1_is}
\end{table}

\begin{table}[htp]
\small
\setlength{\tabcolsep}{4mm}
\centering
\begin{tabular}{c | c c c c}
\toprule
Ret. Modality &  R@1 & R@5 & R@10 \\
\midrule
$T \rightarrow T$ &  25.8 & 65.6 & 79.2 \\
$(I,T) \rightarrow T$ &  28.2 & 67.5 & 80.0 \\
$T \rightarrow (I,T)$ &  26.9 & 67.6 & 79.8 \\
\textbf{$(I,T) \rightarrow (I,T)$} & \textbf{51.3} & \textbf{77.9} & \textbf{82.8} \\
\bottomrule
\end{tabular}
\caption{The ablation study on the impact of different step 2 retrieval modalities on entity-level retrieval results of InfoSeek. Since InfoSeek does not provide annotations for evidence sections, Sec. R@1 is only reported for the results on the E-VQA dataset.}
\label{tab:ablation_step2_modality_is}
\vspace{-4mm}
\end{table}

Table ~\ref{tab:ablation_step1_is} and Table ~\ref{tab:ablation_step2_modality_is}, respectively, present the results of two distinct ablation experiments on InfoSeek, whose results on E-VQA are shown respectively in Tables ~\ref{tab:ablation_step1_evqa} and Tables ~\ref{tab:ablation_step2_modality_evqa}. It is evident that, similar to the results on E-VQA, these findings substantiate the key conclusions of the corresponding ablation studies. This also indirectly demonstrates the soundness and generalizability of our framework design.

\section{Case Study}
\label{sec:case study}
\begin{figure*}[t]
\begin{minipage}[b]{0.325\linewidth}
\scriptsize{\textbf{Q}: In which part of the world does this animal live?\vspace{0.018cm}}\\
\begin{minipage}{0.443\linewidth}
\includegraphics[width=1.\linewidth]{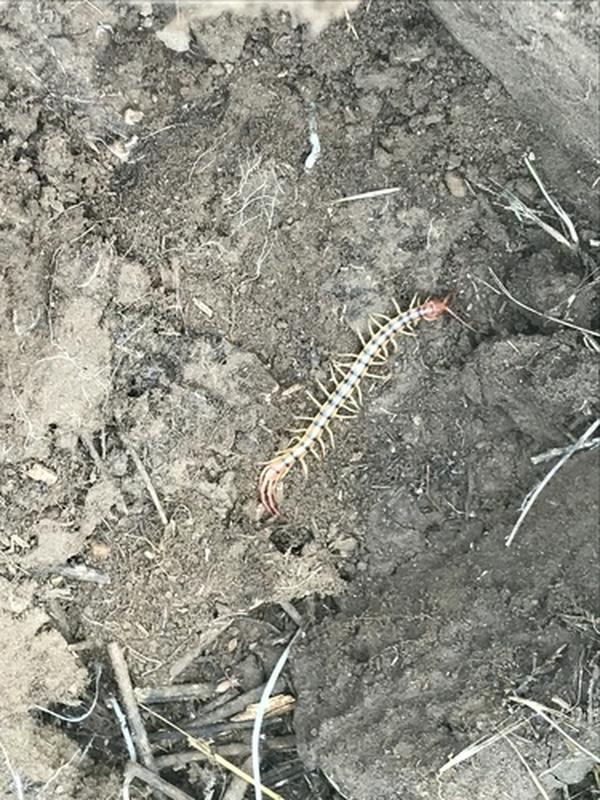}
\end{minipage}
\hfill
\begin{minipage}{0.53\linewidth}
\scriptsize{
\textbf{GPT4-o}:\\
The animal, a centipede, can be found in various parts of the world, particularly in tropical and subtropical regions\textcolor{red}{\XSolidBrush} \\
\textbf{Ours}:\\
Southwestern United States and northern Mexico\textcolor[HTML]{00b050}{\Checkmark}
}
\end{minipage}
\end{minipage}
\hspace{0.02cm}
\begin{minipage}[b]{0.325\linewidth}
\scriptsize{\textbf{Q}: Who is the current curator of this museum?\vspace{0.05cm}}\\
\begin{minipage}{0.443\linewidth}
\includegraphics[width=1.\linewidth]{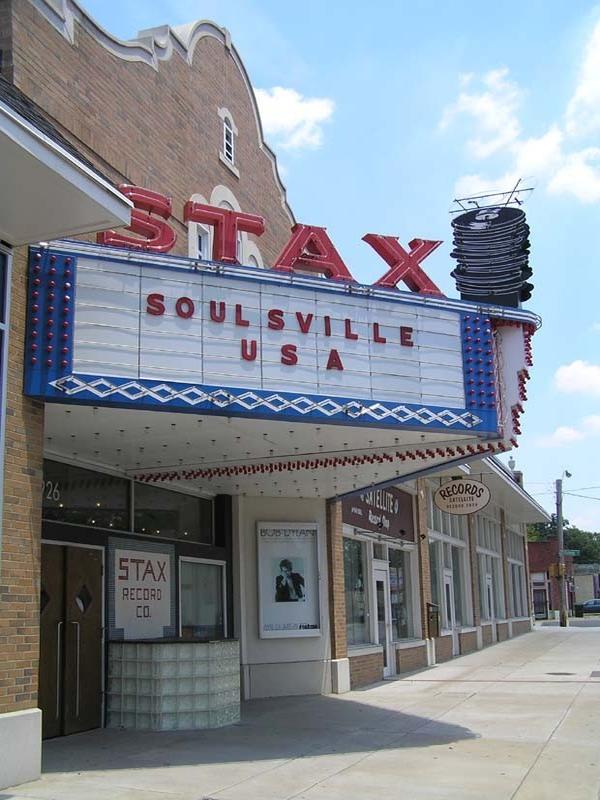}
\end{minipage}
\hfill
\begin{minipage}{0.53\linewidth}
\scriptsize{
\textbf{GPT4-o}:\\
I apologize, but I cannot provide the name of the current curator of the Stax Museum of American Soul Music based solely on the image \textcolor{red}{\XSolidBrush} \\
\textbf{Ours:}\\
Soulsville Foundation \textcolor[HTML]{00b050}{\Checkmark}
}
\end{minipage}
\end{minipage}
\hspace{0.02cm}
\begin{minipage}[b]{0.325\linewidth}
\scriptsize{\textbf{Q}: How many meters tall does this plant grow to?\vspace{0.05cm}}\\
\begin{minipage}{0.443\linewidth}
\includegraphics[width=1.\linewidth]{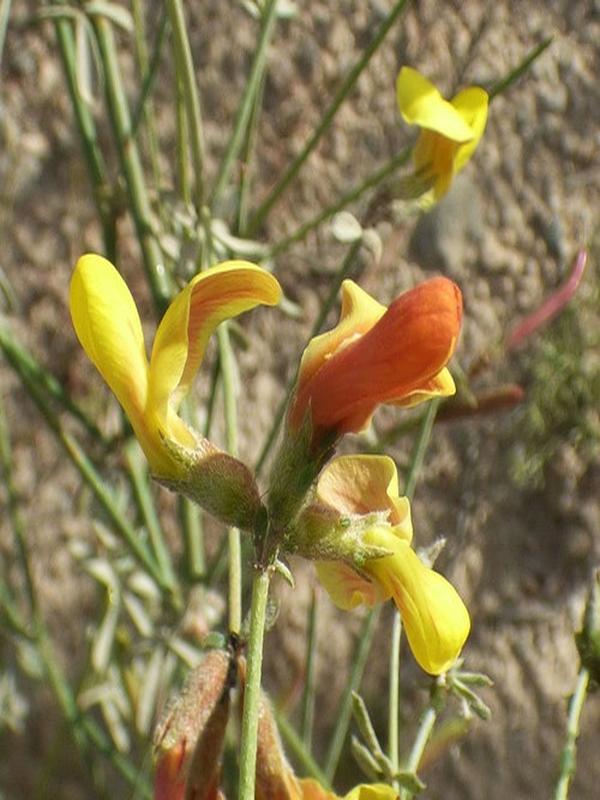}
\end{minipage}
\hfill
\begin{minipage}{0.53\linewidth}
\scriptsize{
\textbf{GPT4-o}:\\
0.1 to 0.5 meters \textcolor{red}{\XSolidBrush} \\
\textbf{Ours:}\\
0.5-1.5 \textcolor[HTML]{00b050}{\Checkmark}
}
\end{minipage}
\end{minipage}
\vspace{0.08cm}

\begin{minipage}[b]{0.325\linewidth}
\scriptsize{\textbf{Q}: What is the source that produces this plant?\vspace{0.11cm}}\\
\begin{minipage}{0.443\linewidth}
\includegraphics[width=1.\linewidth]{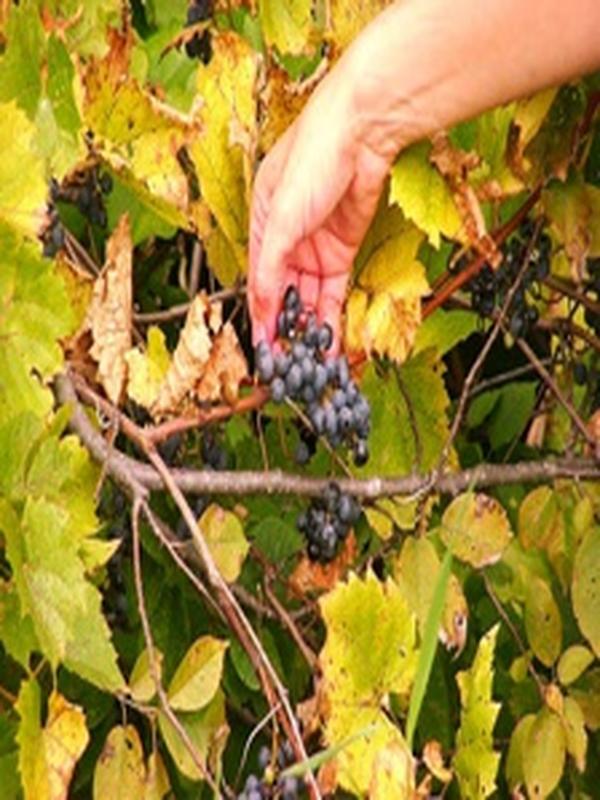}
\end{minipage}
\hfill
\begin{minipage}{0.53\linewidth}
\scriptsize{
\textbf{GPT4-o}:\\
Vitis species (grapevine) \textcolor{red}{\XSolidBrush} \\
\textbf{Ours:}\\
Vitis labrusca \textcolor[HTML]{00b050}{\Checkmark}
}
\end{minipage}
\end{minipage}
\hspace{0.02cm}
\begin{minipage}[b]{0.325\linewidth}
\scriptsize{\textbf{Q}: What is the length of this bridge in metre?\vspace{0.06cm}}\\
\begin{minipage}{0.443\linewidth}
\includegraphics[width=1.\linewidth]{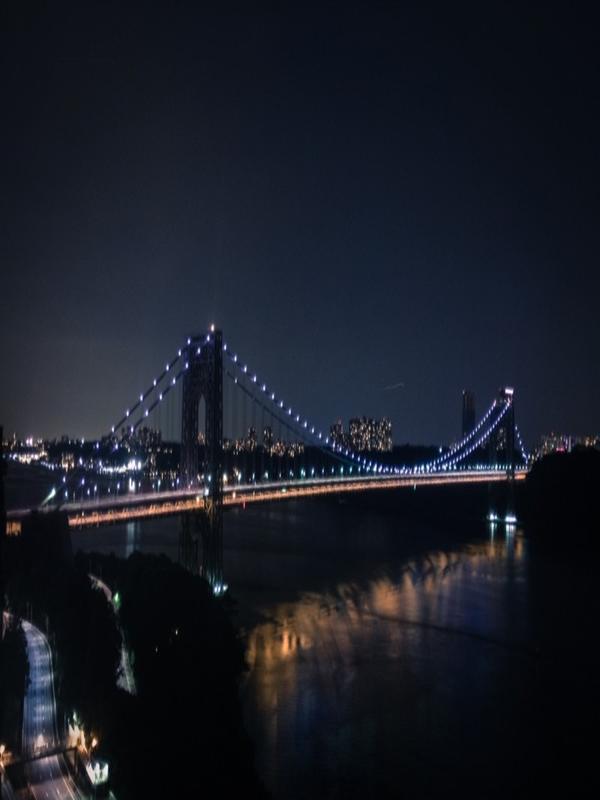}
\end{minipage}
\hfill
\begin{minipage}{0.53\linewidth}
\scriptsize{
\textbf{GPT4-o}:\\
1451 \textcolor{red}{\XSolidBrush} \\
\textbf{Ours:}\\
1450 \textcolor[HTML]{00b050}{\Checkmark}
}
\end{minipage}
\end{minipage}
\hspace{0.02cm}
\begin{minipage}[b]{0.325\linewidth}
\scriptsize{\textbf{Q}: In which year was this item invented or discovered?\vspace{0.05cm}}\\
\begin{minipage}{0.443\linewidth}
\includegraphics[width=1.\linewidth]{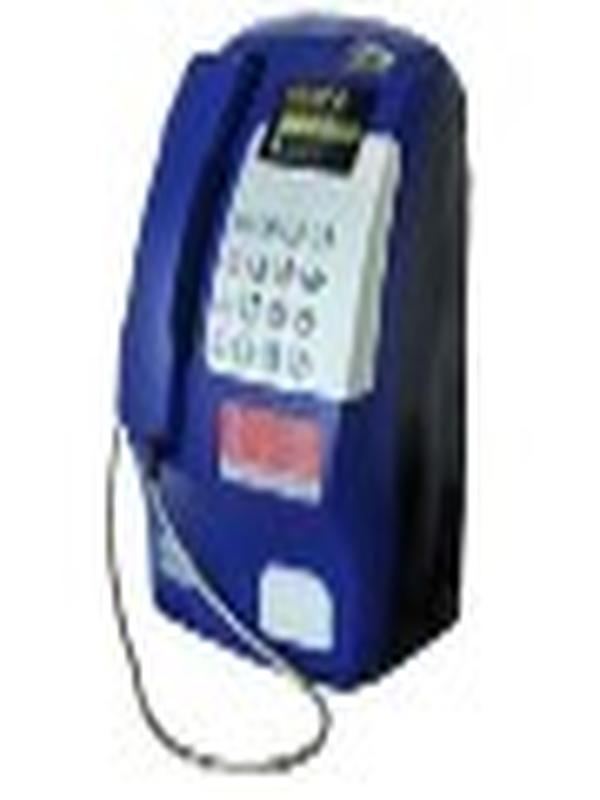}
\end{minipage}
\hfill
\begin{minipage}{0.53\linewidth}
\scriptsize{
\textbf{GPT4-o}:\\
1889 \textcolor{red}{\XSolidBrush} \\
\textbf{Ours:}\\
1876 \textcolor[HTML]{00b050}{\Checkmark}
}
\end{minipage}
\end{minipage}
\vspace{0.08cm}

\begin{minipage}[b]{0.325\linewidth}
\scriptsize{\textbf{Q}: When was this building built?\vspace{0.05cm}}\\
\begin{minipage}{0.443\linewidth}
\includegraphics[width=1.\linewidth]{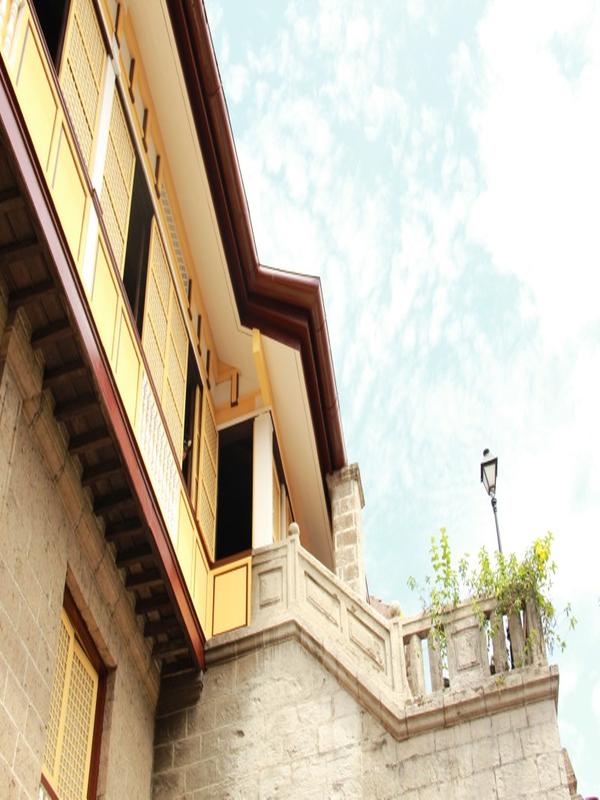}
\end{minipage}
\hfill
\begin{minipage}{0.53\linewidth}
\scriptsize{
\textbf{GPT4-o}:\\
... I can't determine the construction date ... \textcolor{red}{\XSolidBrush} \\
\textbf{Ret. Sec:} \textit{... structure c. \textcolor{red}{1850} ... It was constructed during the \textcolor[HTML]{00b050}{1980s} ...}\\
\textbf{Ours:}\\
1850s \textcolor{red}{\XSolidBrush} \\
\textbf{Ground-truth:}\\
1980s 
}
\end{minipage}
\end{minipage}
\hspace{0.02cm}
\begin{minipage}[b]{0.325\linewidth}
\scriptsize{\textbf{Q}: What is the area in square kilometre occupied by this lake?\vspace{0.05cm}}\\
\begin{minipage}{0.443\linewidth}
\includegraphics[width=1.\linewidth]{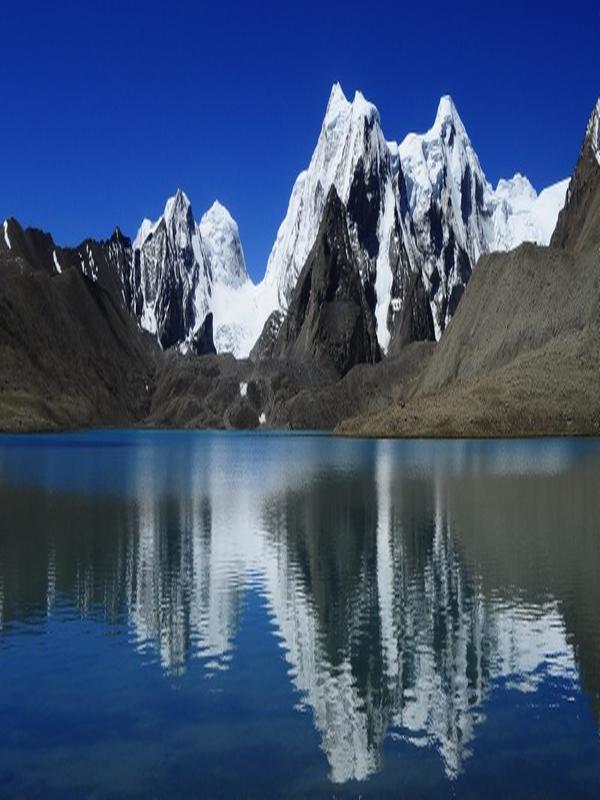}
\end{minipage}
\hfill
\begin{minipage}{0.53\linewidth}
\scriptsize{
\textbf{GPT4-o}:\\
18 \textcolor{red}{\XSolidBrush} \\
\textbf{Ret. Sec:} \textit{... The lake has an area of \textcolor[HTML]{00b050}{118 hectares} ...}\\
\textbf{Ours:}\\
118 \textcolor{red}{\XSolidBrush} \\
\textbf{Ground-truth:}\\
1.18 
}
\end{minipage}
\end{minipage}
\hspace{0.02cm}
\begin{minipage}[b]{0.325\linewidth}
\scriptsize{\textbf{Q}: What is the location of this building?\vspace{0.05cm}}\\
\begin{minipage}{0.443\linewidth}
\includegraphics[width=1.\linewidth]{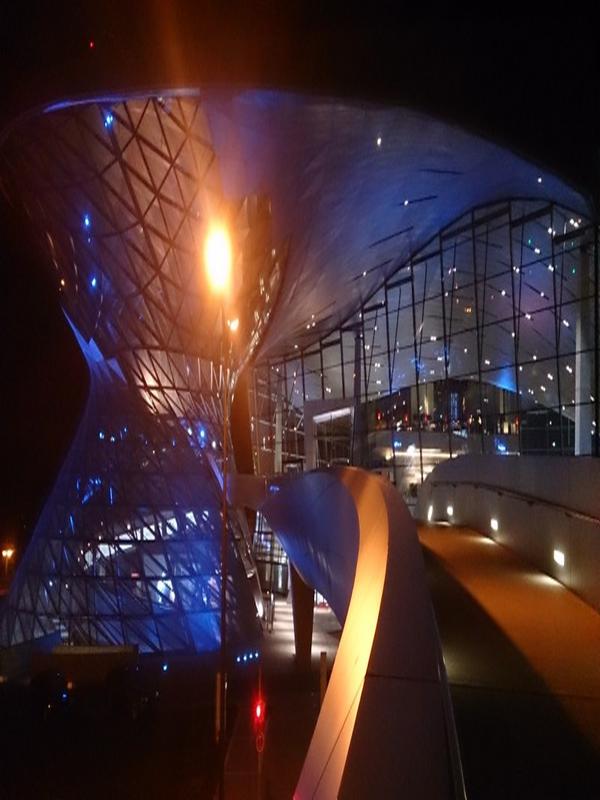}
\end{minipage}
\hfill
\begin{minipage}{0.53\linewidth}
\scriptsize{
\textbf{GPT4-o}:\\
Munich, Germany \textcolor{red}{\XSolidBrush} \\
\textbf{Ret. Sec:} ... located in \textcolor{red}{Munich}'s district \textcolor[HTML]{00b050}{Am Riesenfeld} ...\\
\textbf{Ours:}\\
München \textcolor{red}{\XSolidBrush} \\
\textbf{Ground-truth:}\\
Am Riesenfeld 
}
\end{minipage}
\end{minipage}
\vspace{-0.2cm}
\caption{Qualitative VQA results comparing to GPT4-o. The first row shows results in E-VQA and the second row shows results in InfoSeek. Some failure cases are shown in the third row altogether with ground-truth.}
\label{fig:qualitatives}
\vspace{-0.15cm}
\end{figure*}

To visually assess the performance of our proposed method on the KB-VQA task, we present in the Figure~\ref{fig:qualitatives} the qualitative results on the test dataset for fine-tuned LlaVA-1.5 using OMGM (the best one shown in Table~\ref{tab:vqa_main_results}) and for GPT4-o in zero-shot mode. Evidently, the strong retrieval capability enables the generator to handle a wide range of questions, including those that require precise numerical answers (as shown in the top-right and mid-right examples) and those that involve specialized entity knowledge (as illustrated in the top-left and mid-left examples). In contrast, GPT4-o often fails to identify specific entity information in the images, which leads to either incorrect responses or statements declaring its inability to answer. These examples qualitatively demonstrate the enhanced performance of our method for compact open-source models on the KB-VQA task.

Moreover, the bottom row also displays three failure cases. In these examples, we are generally able to retrieve the corresponding knowledge for the relevant entities. However, several factors may lead to deviations between the generated answers and the true answers: in cases where the lengthy retrieved information contains multiple potential answers, the generator might be unable to accurately extract the specific answer sought by the query (as shown in the bottom-left example); when the output format required by the query differs from that of the retrieved knowledge, the generator may provide an answer in an incorrect format (as seen in the bottom-mid example); and if the relevant content in the retrieved knowledge is nested across multiple layers, the generator might either omit part of the answer or offer a rough response (as illustrated in the bottom-right example). This indicates that although our method achieves excellent retrieval performance, the limitations in the downstream generator’s instruction-following and analytical capabilities still restrict its overall VQA performance.

\begin{table*}[]
\begin{tabular}{p{1\linewidth}}
\hline
\textbf{System}:

You are a Wiki Summary Generator Assistant. Following is some information about you:

\#\# Profile\\
- name: Wiki Summary Generator Assistant \\
- language: English\\
- description: The Wiki Summary Generator Assistant is designed to create concise and informative summaries based on provided Wikipedia content. It extracts key aspects of the entity mentioned in the Wiki article, covering various dimensions such as history, characteristics, significance, appearance and impact.

\#\# Workflows\\
1. Input the provided Wikipedia content into the system.\\
2. Identify the main sections and key information related to the entity.\\
3. Synthesize this information into a well-structured summary.\\
4. Review and refine the summary for clarity, coherence, and completeness before finalizing.

\#\# Rules\\
1. Focus on summarizing key details across multiple aspects (e.g., appearance, features, impact) of the entity.\\
2. Ensure the summary is concise, clear, and free of irrelevant details.\\
3. Retain the original meaning and context of the Wiki content while rephrasing it into a summary.\\
\textbf{User}:\\
Following is the input Wikipedia content: 

\{ \textit{Wikipedia content} \}

Based on the above Wikipedia content, I would like you to generate a summary of the Wikipedia content.
Here is the summary of the Wikipedia content:\\\hline
\end{tabular}
\caption{Prompt template used to instruct the LLM to generate summaries offline for entity knowledge}
\label{tab:summary_generator_prompt}
\end{table*}

\begin{table*}[]
\begin{tabular}{p{1\linewidth}}
\hline
\textbf{System}: \\
\rowcolor[HTML]{FFFFC7}
Answer the encyclopedic question about the given image. Don't mention the visuall content of image in your output. Directly output the answer of the question according to the context. \\
\rowcolor[HTML]{FFCCC9}
You are a helpful assistant for answering encyclopedic questions.\\
\rowcolor[HTML]{CDE6C7}
If the context does not contain the information required to answer the question, you should answer the question using internal model knowledge.\\

\textbf{User}: \\
\rowcolor[HTML]{FFFFC7}
\{ \textit{Query Image} \}\\
\rowcolor[HTML]{CDE6C7}
- Context: \{ \textit{Entity section} \} \\
\rowcolor[HTML]{CDE6C7}
- Question: \{ \textit{Textual question} \} \\
\rowcolor[HTML]{CDE6C7}
The answer is: \\\hline
\end{tabular}
\caption{Prompt used for the VQA process of LLM/MLLM in E-VQA. The \colorbox[HTML]{FFFFC7}{yellow} part is the content only used for MLLM. The \colorbox[HTML]{FFCCC9}{red} part is the content only used for LLM. The \colorbox[HTML]{CDE6C7}{green} part is the content used for both LLM and MLLM.}
\label{tab:evqa_vqa_prompt}
\end{table*}

\begin{table*}[]
\begin{tabular}{p{1\linewidth}}
\hline
\textbf{System}: \\
\rowcolor[HTML]{FFFFC7}
Answer the encyclopedic question about the given image. Don't mention the visuall content of image in your output. Directly output the answer of the question according to the context. \\
\rowcolor[HTML]{FFCCC9}
You are a helpful assistant for answering encyclopedic questions. Do not answer anything else.\\
\rowcolor[HTML]{CDE6C7}
If you need to answer questions about numbers or time, please output the corresponding numerical format directly. If the context does not contain the information required to answer the question, you should answer the question using internal model knowledge.\\
\rowcolor[HTML]{CDE6C7}
There is an example:\\
\rowcolor[HTML]{CDE6C7}
- Context: \# Wiki Article: Dolomites\\
\rowcolor[HTML]{CDE6C7}
\#\# Section Title: Dolomites\\
\rowcolor[HTML]{CDE6C7}
The Dolomites, also known as the Dolomite Mountains, Dolomite Alps or Dolomitic Alps, are a mountain range located in northeastern Italy. The Dolomites are located in the regions of Veneto, Trentino-Alto Adige/Südtirol and Friuli Venezia Giulia, covering an area shared between the provinces of Belluno, Vicenza, Verona, Trentino, South Tyrol, Udine and Pordenone.\\
\rowcolor[HTML]{CDE6C7}
- Question: Which city or region does this mountain locate in?\\
\rowcolor[HTML]{CDE6C7}
Just answer the questions , no explanations needed. Short answer is: Province of Belluno\\

\textbf{User}: \\
\rowcolor[HTML]{FFFFC7}
\{ \textit{Query Image} \}\\
\rowcolor[HTML]{CDE6C7}
- Context: \{ \textit{Entity section} \} \\
\rowcolor[HTML]{CDE6C7}
- Question: \{ \textit{Textual question} \} \\
\rowcolor[HTML]{CDE6C7}
Just answer the questions , no explanations needed. Short answer is: \\\hline
\end{tabular}
\caption{Prompt used for the VQA process of LLM/MLLM in InfoSeek. The \colorbox[HTML]{FFFFC7}{yellow} part is the content only used for MLLM. The \colorbox[HTML]{FFCCC9}{red} part is the content only used for LLM. The \colorbox[HTML]{CDE6C7}{green} part is the content used for both LLM and MLLM.}
\label{tab:infoseek_vqa_prompt}
\end{table*}

\end{document}